\documentstyle[12pt]{article}
\newcommand{\nn}{\nonumber}
\newcommand{\bd}{\begin{document}}
\newcommand{\ed}{\end{document}}
\newcommand{\bc}{\begin{center}}
\newcommand{\ec}{\end{center}}
\newcommand{\be}{\begin{eqnarray}}
\newcommand{\ee}{\end{eqnarray}}
\renewcommand{\thefootnote}{\alph{footnote}}
\newcommand{\se}{\section}
\newcommand{\sse}{\subsection}
\newcommand{\bi}{\bibitem}
\newcommand{\text}{\rm}
\newcommand{\func}{\rm}
\def\figcap{\section*{Figure Captions\markboth
     {FIGURECAPTIONS}{FIGURECAPTIONS}}\list
     {Figure \arabic{enumi}:\hfill}{\settowidth\labelwidth{Figure 999:}
     \leftmargin\labelwidth
     \advance\leftmargin\labelsep\usecounter{enumi}}}
\let\endfigcap\endlist \relax
\begin{document}

\begin{titlepage}

 \vskip 0.5in
 \null
\begin{center}
 \vspace{.15in}
{\Large {\bf Baryonic Rare Decays of $\Lambda_b\to \Lambda \ell^+
\ell^-$}
}\\
\vspace{1.0cm}  \par
 \vskip 2.1em
 {\large
  \begin{tabular}[t]{c}
{\bf Chuan-Hung Chen$^a$ and C.~Q.~Geng$^b$}
\\
\\
       {\sl ${}^a$Department of Physics, National Cheng Kung University}
\\   {\sl  $\ $Tainan, Taiwan,  Republic of China }
\\
\\
{\sl ${}^b$Department of Physics, National Tsing Hua University}
\\  {\sl  $\ $ Hsinchu, Taiwan, Republic of China }
\\
   \end{tabular}}
 \par \vskip 5.3em

 {\Large\bf Abstract}
\end{center}
We present a systematic analysis for the rare baryonic exclusive decays
of $\Lambda_b\to\Lambda l^+ l^-\ (l=e,\mu,\tau )$.
% in the standard and SUSY models.
We study the differential decay rates and the di-lepton
forward-backward, lepton polarization and various CP asymmetries
with a new simple set of form factors inspired by the 
heavy quark effective theory. We show that most of the observables are
insensitive to the non-perturbative QCD effects. To illustrate the
effect of new physics, we discuss our results in an explicit
supersymmetric extension of the standard model, which contains new CP
violating phases and therefore induces sizable CP violating
asymmetries.

\end{titlepage}

\newpage

\section{Introduction}

A priority in current particle physics research is to determine
the parameters of the Cabibbo-Kobayashi-Maskawa (CKM) matrix elements
\cite{ckm} in the standard model (SM).
 Due to the CLEO measurement of the radiative $b \to s\gamma$ decay
\cite{cleo}, some interest has
been focused on the rare decays related to $b\to s l^+l^-$ induced by
the flavor changing neutral currents (FCNCs).
In the SM, these rare decays occur at the loop level and depend
on the CKM elements.
In the literature, most of studies have been concentrated on the
corresponding exclusive rare B-meson decays such as $B\to K^{(*)}l^+l^-$
\cite{Bmeson}. However, these exclusive modes contain several unknown
hadronic form factors, which cannot be measured in the present B-meson
facilities unlike the kaon cases.
Recently, we have examined the exclusive rare baryonic decays of
$\Lambda_b\to\Lambda l\,\bar{l}\ (l=\nu,e,\mu,\tau)$
\cite{chen0,chen1,chen2} and found that some of physical quantities
are insensitive to the hadronic uncertainties.

In this paper, we give a systematic study on the baryonic decays
of $\Lambda_b\to\Lambda l^+l^-$.
We will explore various possible CP even and odd asymmetries
to show how the hadronic unknown parameters are factored out in most of
cases. To illustrate CP violating effect, we will also discuss an 
explicit CP violating model with SUSY.

The paper is organized as follows. In Sec.~2, we give the
effective Hamiltonian for the decays of $\Lambda_b \to \Lambda l
\bar{l} $ and the most general form factors in the $\Lambda_b \to
\Lambda$ transition. In Sec.~3, we derive the general forms of the
differential decay rates. In Sec.~4, we study the di-lepton
forward-backward, lepton polarization and various CP violating
asymmetries. We perform our numerical analysis in Sec.~5. We
present our conclusions in Sec.~6.

\section{Effective Hamiltonian and form factors}

In the SM, the effective Hamiltonian for $b\rightarrow
sl^{+}l^{-}$ is given by

\begin{equation}
{\cal H}=-4\frac{G_{F}}{\sqrt{2}}V_{tb}V_{ts}^{*}\sum_{i=1}^{10}C_{i}\left(
\mu \right) O_{i}\left( \mu \right) \,  \label{Ham}
\end{equation}
where 
the expressions of the renormalized Wilson coefficients
$C_{i}(\mu )$ and operators $O_{i}(\mu )$ 
 can be found in Ref. \cite{Buras}. 
 From Eq.
(\ref{Ham}), the free quark decay amplitude 
 is written as
\begin{eqnarray}
{\cal M}\left( b\rightarrow sl^{+}l^{-}\right) &=&\frac{G_{F}\alpha _{em}}{%
\sqrt{2}\pi }V_{tb}V_{ts}^{*}\left[ \bar{s}\left( C_{9}^{eff}\left( \mu
\right) \gamma _{\mu }P_{L}-\frac{2m_{b}}{q^{2}}C_{7}^{eff}\left( \mu
\right) i\sigma _{\mu \nu }q^{\nu }P_{R}\right) b\;\bar{l}\gamma ^{\mu
}l\right.  \nonumber \\
&&\left. +\bar{s}C_{10}\gamma _{\mu }P_{L}b\;\bar{l}\gamma ^{\mu }\gamma
_{5}l\right]  \label{Am}
\end{eqnarray}
with $P_{L(R)}=(1\mp \gamma _{5})/2$. We note that in Eq.
(\ref{Am}), only the term associated with the Wilson coefficient
$C_{10}$ is independent of the $\mu $ scale. Besides the
short-distance (SD)  contributions, the long-distance (LD) ones
such as that from the $c\bar{c}$ resonant states of $\Psi ,\Psi
^{\prime }...etc$ are also important for the decay rate. It is
known that for the LD effects in the B-meson decays
\cite{DTP,LMS,AMM,OT,KS,Geng1}, both the factorization
assumption (FA) and the vector meson dominance (VMD) approximation
have been used. In baryonic decays, we assume that the
parametrization of LD contributions is the same as that in the 
B-meson decays. Hence, we may include the resonant effect (RE) by
absorbing it to the corresponding Wilson coefficient. In this
paper as a more complete analysis we also include the LD contributions
to the decay of $b\rightarrow s\gamma $, induced by the
nonfactorizable effects \cite{Soare,MNS}. The effective Wilson
coefficients of $C_{9}^{eff}$ and $C_{7}^{eff}$ can be expressed as the
standard form
\begin{eqnarray}
C_{9}^{eff}\left( \mu \right) &=&C_{9}\left( \mu \right) +Y\left(
z,s^{\prime }\right)\,,  \label{ceff9} \\
C_{7}^{eff}\left( \mu \right) &=&C_{7}\left( \mu \right) +C_{7}^{\prime
}\left( \mu ,q^{2}\right)\,,
\label{ceff7}
\end{eqnarray}
where
\begin{eqnarray}
Y\left( z,s^{\prime }\right) &=&\left( h(z,s^{\prime })+\frac{3}{\alpha
_{em}^{2}}\sum_{j=\Psi ,\Psi ^{\prime }}k_{j}\frac{\pi \Gamma \left(
j\rightarrow l^{+}l^{-}\right) M_{j}}{q^{2}-M_{j}^{2}+iM_{j}\Gamma _{j}}%
\right)
\nn \\
&&-\frac{1}{2}h(1,s^{\prime })\left( 4C_{3}+4C_{4}+3C_{5}+C_{6}\right) -%
\frac{1}{2}h(0,s^{\prime })\left( C_{3}+3C_{4}\right)\,,
\nn\\
C_{7}^{\prime }\left( \mu ,q^{2}\right) &=&C_{b\rightarrow s\gamma }^{\prime
}\left( \mu \right) +\omega \left( h(z,s^{\prime })+\frac{3}{\alpha _{em}^{2}%
}\sum_{j=\Psi ,\Psi ^{\prime }}k_{j}\frac{\pi \Gamma \left( j\rightarrow
l^{+}l^{-}\right) M_{j}}{q^{2}-M_{j}^{2}+iM_{j}\Gamma _{j}}\right)\,,
\end{eqnarray}
with
\begin{eqnarray}
h(z,s^{\prime }) &=&-\frac{8}{9}\ln z+\frac{8}{27}+\frac{4}{9}x-\frac{2}{9}%
\left( 2+x\right) \left| 1-x\right| ^{1/2}  \nonumber \\
&&\times \left\{
\begin{array}{l}
\ln \left| \frac{\sqrt{1-x}+1}{\sqrt{1-x}-1}\right| -i\pi \quad {\rm {for}}%
{\rm {\ }x\equiv 4z^{2}/s^{\prime }<1} \\
2\arctan \frac{1}{\sqrt{x-1}}\qquad {\rm {for}}{\rm {\ }x\equiv
4z^{2}/s^{\prime }>1}
\end{array}
\right. \,, \nn\\
C_{b\rightarrow s\gamma }^{\prime } &=&i\alpha _{s}\left[ \frac{2}{9}\eta
^{14/23}\left( G_{1}\left( x_{t}\right) -0.1687\right) -0.03C_{2}\left( \mu
\right) \right] ,  \nonumber \\
G_{1}\left( x\right) &=&\frac{x\left( x^{2}-5x-2\right) }{8\left( x-1\right)
^{3}}+\frac{3x^{2}\ln x}{4\left( x-1\right) ^{4}} \,.
\label{hf}
\end{eqnarray}
Here $Y(z,s^{\prime })$ combines the one-loop matrix elements and
the LD contributions of operators $O_{1}$ - $O_{6}$,
$C_{b\rightarrow s\gamma }^{\prime }$ is the absorptive part of
$b\rightarrow s\gamma $  \cite{SAI} with neglecting
the small contribution from $V_{ub}V_{us}^{*}$, $%
z=m_{c}/m_{b},$ $s^{\prime }=q^{2}/m_{b}^{2},$ $\eta =\alpha _{s}\left(
m_{W}\right) /\alpha _{s}\left( \mu \right)$, $%
x_{t}=m_{t}^{2}/m_{W}^{2} $,
  $M_{j}\ (\Gamma
_{j})$ are the masses (widths) of intermediate states, $\left|
\omega \right| \leq 0.15$  describing the nonfactorizable
contributions to $b\rightarrow s\gamma $ decay at $q^{2}=0$
\cite{Soare,MNS}, and the factors $k_{j}$ are phenomenological
parameters for compensating the approximations of the FA and VMD
and reproducing the correct branching ratios of $B(\Lambda _{b}\to
\Lambda J/\Psi \to \Lambda l^{+}l^{-})=B(\Lambda _{b}\to \Lambda J/\Psi
)\times B(J/\Psi \to l^{+}l^{-})$ when we study the $\Lambda_b$
decays. We note that by taking $k_{\Psi }\simeq -1/(3C_{1}+C_{2})$
and $B(\Lambda _{b}\to \Lambda J/\Psi )=\left( 4.7\pm 2.8\right) \times
10^{-4}$, the $k_{j}$ factors in the $\Lambda_b$ case are almost
the same as that in the B-meson one \cite{chen1}.
The Wilson coefficients (WCs) at the scale of $\mu
\sim m_{b}\sim 4.8$ GeV  are shown in Table
\ref{wc}.
\begin{table}[h]
\caption{ Wilson coefficients for $m_{t}=170$ GeV, $\mu =4.8$ GeV.}
\label{wc}
\begin{center}
\[
\begin{tabular}{|c|c|c|c|c|c|}
\hline
$WC$ & $C_{1}$ & $C_{2}$ & $C_{3}$ & $C_{4}$ & $C_{5}$ \\ \cline{2-6}
& $-0.226$ & $1.096$ & $0.01$ & $-0.024$ & $0.007$ \\ \hline
$WC$ & $C_{6}$ & $C_{7}$ & $C_{8}$ & $C_{9}$ & $C_{10}$ \\ \cline{2-6}
& $-0.028$ & $-0.305$ & $-0.15$ & $4.186$ & $-4.559$ \\ \hline
\end{tabular}
\]
\end{center}
\end{table}

 Using the form factors given in Appendix A, we write 
%From Eqs. (\ref{vc}), (\ref{ac}), (\ref{tcq}), and (\ref{atcq}),
the amplitude of $\Lambda _{b}\rightarrow
\Lambda l^{+}l^{-}$ as

\begin{equation}
{\cal M}\left( \Lambda _{b}\rightarrow \Lambda l^{+}l^{-}\right) =\frac{%
G_{F}\alpha _{em}}{\sqrt{2}\pi }V_{tb}V_{ts}^{*}\left\{ H_{1}^{\mu }L_{\mu
}+H_{2}^{\mu }L_{\mu }^{5}\right\}  \label{da}
\end{equation}
where
\begin{eqnarray}
H_{1}^{\mu } &=&\bar{\Lambda}\gamma ^{\mu }\left(
A_{1}P_{R}+B_{1}P_{L}\right) \Lambda _{b}+\bar{\Lambda}i\sigma ^{\mu \nu
}q_{\nu }\left( A_{2}P_{R}+B_{2}P_{L}\right) \Lambda _{b}\,,
\nn \\
H_{2}^{\mu } &=&\bar{\Lambda}\gamma ^{\mu }\left(
D_{1}P_{R}+E_{1}P_{L}\right) \Lambda _{b}+\bar{\Lambda}i\sigma ^{\mu \nu
}q_{\nu }\left( D_{2}P_{R}+E_{2}P_{L}\right) \Lambda _{b} \\
&&+q^{\mu }\bar{\Lambda}\left( D_{3}P_{R}+E_{3}P_{L}\right) \Lambda
_{b}\,, \nn\\
L_{\mu } &=&\ \bar{l}\gamma _{\mu }l\,,\nn \\
L_{\mu }^{5} &=&\bar{l}\gamma _{\mu }\gamma _{5}l
\end{eqnarray}
with
\begin{eqnarray}
A_{i} &=&C_{9}^{eff}\frac{f_{i}-g_{i}}{2}-\frac{2m_{b}}{q^{2}}C_{7}^{eff}%
\frac{f_{i}^{T}{}+g_{i}^{T}}{2},  \nonumber \\
B_{i} &=&C_{9}^{eff}\frac{f_{i}+g_{i}}{2}-\frac{2m_{b}}{q^{2}}C_{7}^{eff}%
\frac{f_{i}^{T}-g_{i}^{T}}{2},  \nonumber \\
D_{i} &=&C_{10}\frac{f_{i}-g_{i}}{2},  \nonumber \\
E_{i} &=&C_{10}\frac{f_{i}+g_{i}}{2}.  \label{coeffs}
\end{eqnarray}
and $i=1,2,3$.

The processes for the heavy to light baryonic decays such as those
with $\Lambda _{b}\rightarrow \Lambda $ have been studied based on
the heavy quark effective theory (HQET) in Ref. \cite{MR}  and it
is found that
\be
\left\langle \Lambda (p_{\Lambda })\right| \bar{s}\Gamma b\left| \Lambda
_{b}(p_{\Lambda _{b}})\right\rangle =\bar{u}_{\Lambda }\left( F_{1}(q^{2})+%
\not{v}F_{2}(q^{2})\right) \Gamma u_{\Lambda _{b}} \ee where
$\Gamma $ denotes the Dirac matrix, $v=p_{\Lambda _{b}}/M_{\Lambda
_{b}} $ is the four-velocity of $\Lambda _{b}$, $q=p_{\Lambda
_{b}}-p_{\Lambda }$ is the momentum transfer, and $F_{1,2}$ are
the form factors. Clearly, there are only two independent form
factors $F_{1,2}$ in the HQET. Comparing with the general forms of the
form factors in Appendix A,
%Eqs. (\ref{vc}), (\ref{ac}), (\ref{tcq}), and ( {\ref{atcq}), 
we get the relations among the form factors as follows:
\begin{eqnarray}
g_{1} &=&f_{1}=f_{2}^{T}=g_{2}^{T}=F_{1}+\sqrt{r}F_{2},  \nonumber \\
g_{2} &=&f_{2}=g_{3}=f_{3}=g_{T}^{V}=f_{T}^{V}=\frac{F_{2}}{M_{\Lambda _{b}}}%
,  \nonumber \\
g_{T}^{S} &=&f_{T}^{S}=0,  \nonumber \\
g_{1}^{T} &=&f_{1}^{T}=\frac{F_{2}}{M_{\Lambda _{b}}}q^{2},  \nonumber \\
g_{3}^{T} &=&\frac{F_{2}}{M_{\Lambda _{b}}}\left( M_{\Lambda
_{b}}+M_{\Lambda }\right) ,\quad f_{3}^{T}=-\frac{F_{2}}{M_{\Lambda _{b}}}%
\left( M_{\Lambda _{b}}-M_{\Lambda }\right) ,  \label{ffr}
\end{eqnarray}
where $r=M_{\Lambda }^{2}/M_{\Lambda _{b}}^{2}$.
 From the CLEO result of
 $R=-0.25\pm 0.14\pm 0.08$ \cite{CLEO}, we know that $|F_2|<|F_1|$.
Due to Eq. (\ref{ffr}), only $f_{1}\,(g_{1}) $ and $f_{2}^{T}\,(
g_{2}^{T}) $ are proportional to $F_1$ and therefore, they are large,
whereas all the others are small since they are related to the small form
factor $F_2$. 
Furthermore, 
 from Eq. (\ref{coeffs}), we
 find that $\{f^{T}\}$ and $\{g^{T}\}$ are associated
with $C_{7}$ which is about one order of the magnitude smaller than
$C_{9}$ and $C_{10}$ so that their effects to
 the deviation of the results in the HQET are small.
Hence, with the information of the HQET, we can make a good approximation
for the general form factors of
transition matrix elements given in Eqs. (\ref{da})  and (\ref{coeffs}).
Altogether,
we have the following relations:
\begin{eqnarray}
\bar{f} &\equiv &\frac{f_{1}+g_{1}}{2},\quad
\frac{f_{2}^{T}+g_{2}^{T}}{f_{1}+g_{1}%
}\simeq 1  \nonumber \\
\frac{f_{1}-g_{1}}{f_{1}+g_{1}} &\simeq &\delta ,\quad
\frac{g_{2}}{f_{2}}%
\simeq \frac{g_{1}^{T}}{f_{1}^{T}}\simeq \frac{g_{2}^{T}}{f_{2}^{T}}%
\simeq 1,  \nonumber \\
\frac{f_{1}^{T}+g_{1}^{T}}{f_{1}+g_{1}}\frac{1}{q^{2}} &\simeq &\frac{%
f_{2}+g_{2}}{f_{1}+g_{1}}\,.
\label{appff}
\end{eqnarray}
In the HQET, it is easy to show that
\be
 \delta & =& 0\,,\
\rho\equiv  M_{\Lambda_b} \left({f_2+g_2\over f_1+g_1}\right)
={F_2\over F_1+\sqrt{r}F_2}\,.
\label{HQETff}
\ee

\section{Differential decays rates}

In this section we first present the formulas by including the lepton mass
for the double differential decay rates with respect to the angle of
the lepton and the invariant mass of the di-lepton. In the following we
only show the results of  the SM with the form factors in Eq.
(\ref{appff}). The general ones with including right-handed
couplings are presented in Appendix B.

Introducing dimensionless variables  of $t=p_{\Lambda _{b}}\cdot
p_{\Lambda
}/M_{\Lambda _{b}}^{2},$ $r=M_{\Lambda }^{2}/M_{\Lambda _{b}}^{2},$ $\hat{m}%
_{l}=m_{l}/M_{\Lambda _{b}},$ $\hat{m}_{b}=m_{b}/M_{\Lambda _{b}},$ and $%
s=q^{2}/M_{\Lambda _{b}}^{2}$, the double partial differential decay rates
for $\Lambda _{b}\rightarrow \Lambda \ l^{+}\ l^{-}$ $\left( l=e,\ \mu ,\
\tau \right) $ can be written as
\be
\frac{d^{2}\Gamma }{dsd\hat{z}} &=&\frac{G_{F}^{2}\alpha _{em}^{2}\lambda
_{t}^{2}}{768\pi ^{5}}M_{\Lambda _{b}}^{5}\sqrt{\phi \left( s\right) }\sqrt{%
1-\frac{4m_{l}^{2}}{q^{2}}}\bar{f}^{2}R_{\Lambda _{b}}\left( s,\hat{z}%
\right) ,  \label{dorate}
\ee
where
\be
R_{\Lambda _{b}}\left( s,\hat{z}\right) &=&I_{0}\left(
s,\hat{z}\right) +%
\hat{z}I_{1}\left( s,\hat{z}\right) +\hat{z}^{2}I_{2}\left( s,\hat{z}\right)
\ee
and
\begin{eqnarray}
I_{0}\left( s,\hat{z}\right) &=&-6\sqrt{r}s\left[ -2\hat{m}_{b}\rho \left(
1+2\frac{m_{l}^{2}}{q^{2}}\right) {\rm {Re}C_{9}^{eff}C_{7}^{eff*}}\right.
\nonumber \\
&&\left. +\delta \left( \left( 1+2\frac{m_{l}^{2}}{q^{2}}\right) \left|
C_{9}^{eff}\right| ^{2}+\left( 1-6\frac{m_{l}^{2}}{q^{2}}\right) \left|
C_{10}\right| ^{2}\right) \right]  \nonumber \\
&&+\frac{3}{4}\left( \left( 1-r\right) ^{2}-s^{2}\right) \left[ \left( 2\hat{%
m}_{b}\rho \right) ^{2}\left| C_{7}^{eff}\right| ^{2}+\left|
C_{9}^{eff}\right| ^{2}+\left| C_{10}\right| ^{2}\right]  \nonumber \\
&&+6\hat{m}_{l}^{2}t\left[ \left( 2\hat{m}_{b}\rho \right) ^{2}\left|
C_{7}^{eff}\right| ^{2}+\left| C_{9}^{eff}\right| ^{2}-\left| C_{10}\right|
^{2}\right]  \nonumber \\
&&+6\sqrt{r}\left( 1-t\right) \left\{ 4\left( 1+2\frac{m_{l}^{2}}{q^{2}}%
\right) \hat{m}_{b}^{2}\rho \left| C_{7}^{eff}\right| ^{2}\right.  \nonumber
\\
&&\left. +\rho s\left[ \left( 1+2\frac{m_{l}^{2}}{q^{2}}\right) \left|
C_{9}^{eff}\right| ^{2}+\left( 1-2\frac{m_{l}^{2}}{q^{2}}\right) \left|
C_{10}\right| ^{2}\right] \right\}  \nonumber \\
&&+12\left( 1+2\frac{m_{l}^{2}}{q^{2}}\right) \hat{m}_{b}\left( t-r\right)
\left( 1+s\rho ^{2}\right) {\rm {Re}C_{9}^{eff}C_{7}^{eff*}}  \nonumber \\
&&+12\left( 1+2\frac{m_{l}^{2}}{q^{2}}\right) \hat{m}_{b}\sqrt{r}s\rho {\rm {%
Re}C_{9}^{eff}C_{7}^{eff*}}  \nonumber \\
&&-6\left[ s\left( 1-t\right) \left( t-r\right) -\frac{1}{8}\left( \left(
1-r\right) ^{2}-s^{2}\right) \right]  \nonumber \\
&&\times \left[ \frac{4\hat{m}_{b}^{2}}{s}\left| C_{7}^{eff}\right|
^{2}+s\rho ^{2}\left( \left| C_{9}^{eff}\right| ^{2}+\left| C_{10}\right|
^{2}\right) \right]  \nonumber \\
&&-6\hat{m}_{l}^{2}\left( 2r-\left( 1+r\right) t\right) \left[ \left( \frac{2%
\hat{m}_{b}}{s}\right) ^{2}\left| C_{7}^{eff}\right| ^{2}+\rho ^{2}\left(
\left| C_{9}^{eff}\right| ^{2}-\left| C_{10}\right| ^{2}\right) \right] ,
\label{I0} \\
I_{1}\left( s,\hat{z}\right) &=&3\sqrt{1-\frac{4m_{l}^{2}}{q^{2}}}\phi
\left( s\right) \left\{ s\left[ 1-2\sqrt{r}\rho -\left( 1-r\right) \rho
^{2}\right] {\rm {Re}C_{9}^{eff}C_{10}^{*}}\right.  \nonumber \\
&&\left. +2\hat{m}_{b}\left( 1-s\rho ^{2}\right) {\rm {Re}%
C_{7}^{eff}C_{10}^{*}}\right\} ,  \label{I1} \\
I_{2}\left( s,\hat{z}\right) &=&-\frac{3}{4}\phi \left( s\right) \left( 1-4%
\frac{m_{l}^{2}}{q^{2}}\right) \left[ \left( 2\hat{m}_{b}\rho \right)
^{2}\left| C_{7}^{eff}\right| ^{2}+\left| C_{9}^{eff}\right| ^{2}+\left|
C_{10}\right| ^{2}\right]  \nonumber \\
&&+\frac{3}{4}\phi \left( s\right) \left( 1-4\frac{m_{l}^{2}}{q^{2}}\right)
\left[ \frac{4\hat{m}_{b}^{2}}{s}\left| C_{7}^{eff}\right| ^{2}+s\rho
^{2}\left( \left| C_{9}^{eff}\right| ^{2}+\left| C_{10}\right| ^{2}\right)
\right] ,  \label{I2}
\end{eqnarray}
with $\hat{z}=\hat{p}_{B}\cdot \hat{p}_{l^{+}}$ being the angle between
the momenta of $\Lambda _{b}$ and $l^+$ in the di-lepton
invariant mass frame and $\phi \left( s\right) =\left( 1-r\right)
^{2}-2s\left(1+r\right) +s^{2}$.
Here, for the simplicity, we have not
displayed the dependence of the $\mu $ scale in effective Wilson
coefficients.
We note that the main nonperturbative QCD effect from $\bar{f}$ has been
factored out in Eq. (\ref{dorate}).
The function $R_{\Lambda
_{b}}(s,\hat{z})$ is only related to the two parameters of
$\delta $ and $\rho $ which become one in the HQET.
Since $\rho$
 is the ratio of form factors and insensitive to the QCD models,
 the QCD effects in the baryonic di-lepton decays
 are clearly less significant.
Therefore, these decay modes
  are good physical observable to test the SM.

After integrating the angular dependence, the invariant mass distributions
as function of $s$ are given by

\begin{equation}
\frac{d\Gamma \left( \Lambda _{b}\rightarrow \Lambda l^{+}l^{-}\right) }{ds}=%
\frac{G_{F}^{2}\alpha _{em}^{2}\lambda _{t}^{2}}{384\pi ^{5}}M_{\Lambda
_{b}}^{5}\sqrt{\phi \left( s\right) }\sqrt{1-\frac{4m_{l}^{2}}{q^{2}}}\bar{f}%
^{2}R_{\Lambda _{b}}\left( s\right) ,  \label{rate}
\end{equation}
where
\be
R_{\Lambda _{b}}\left( s\right) =\Gamma _{1}\left( s\right) +\Gamma
_{2}\left( s\right) +\Gamma _{3}\left( s\right)
\ee
with
\begin{eqnarray}
\Gamma _{1}\left( s\right) &=&-6\sqrt{r}s\left[ -2\hat{m}_{b}\rho \left( 1+2%
\frac{m_{l}^{2}}{q^{2}}\right) {\rm {Re}C_{9}^{eff}C_{7}^{eff*}}\right.
\nonumber \\
&&\left. +\delta \left( \left( 1+2\frac{m_{l}^{2}}{q^{2}}\right) \left|
C_{9}^{eff}\right| ^{2}+\left( 1-6\frac{m_{l}^{2}}{q^{2}}\right) \left|
C_{10}\right| ^{2}\right) \right]  \nonumber \\
&&+\left[ -2r\left( 1+2\frac{m_{l}^{2}}{q^{2}}\right) -4t^{2}\left( 1-\frac{%
m_{l}^{2}}{q^{2}}\right) +3\left( 1+r\right) t\right]  \nonumber \\
&&\times \left[ \left( 2\hat{m}_{b}\rho \right) ^{2}\left|
C_{7}^{eff}\right| ^{2}+\left| C_{9}^{eff}\right| ^{2}+\left| C_{10}\right|
^{2}\right]  \nonumber \\
&&+6\hat{m}_{l}^{2}t\left[ \left( 2\hat{m}_{b}\rho \right) ^{2}\left|
C_{7}^{eff}\right| ^{2}+\left| C_{9}^{eff}\right| ^{2}-\left| C_{10}\right|
^{2}\right] ,  \label{rate1} \\
\Gamma _{2}\left( s\right) &=&6\sqrt{r}\left( 1-t\right) \left\{ 4\left( 1+2%
\frac{m_{l}^{2}}{q^{2}}\right) \hat{m}_{b}^{2}\rho \left| C_{7}^{eff}\right|
^{2}\right.  \nonumber \\
&&\left. +\rho s\left[ \left( 1+2\frac{m_{l}^{2}}{q^{2}}\right) \left|
C_{9}^{eff}\right| ^{2}+\left( 1-2\frac{m_{l}^{2}}{q^{2}}\right) \left|
C_{10}\right| ^{2}\right] \right\}  \nonumber \\
&&+12\left( 1+2\frac{m_{l}^{2}}{q^{2}}\right) \hat{m}_{b}\left( t-r\right)
\left( 1+s\rho ^{2}\right) {\rm {Re}C_{9}^{eff}C_{7}^{eff*}},  \label{rate2}
\\
\Gamma _{3}\left( s\right) &=&12\left( 1+2\frac{m_{l}^{2}}{q^{2}}\right)
\hat{m}_{b}\sqrt{r}s\rho {\rm {Re}C_{9}^{eff}C_{7}^{eff*}}  \nonumber \\
&&-\left[ 2t^{2}\left( 1+2\frac{m_{l}^{2}}{q^{2}}\right) +4r\left( 1-\frac{%
m_{l}^{2}}{q^{2}}\right) -3\left( 1+r\right) t\right]  \nonumber \\
&&\times \left[ \frac{4\hat{m}_{b}^{2}}{s}\left| C_{7}\right| ^{2}+s\rho
^{2}\left( \left| C_{9}^{eff}\right| ^{2}+\left| C_{10}\right| ^{2}\right)
\right]  \nonumber \\
&&-6\hat{m}_{l}^{2}\left( 2r-\left( 1+r\right) t\right) \left[ \left( \frac{2%
\hat{m}_{b}}{s}\right) ^{2}\left| C_{7}^{{\rm eff}}\right| ^{2}+\rho
^{2}\left( \left| C_{9}^{{\rm eff}}\right| ^{2}-\left| C_{10}\right|
^{2}\right) \right]\,.  \label{rate3}
\end{eqnarray}
The limits for $s$ are given by
\be
4\hat{m}_{l}^{2}\leq s\leq \left( 1-\sqrt{r}\right) ^{2}.
\ee
 From Eqs. (\ref{rate1})-(\ref{rate3}), we see
that $\rho $ appears either as $\sqrt{r}\rho $ or $\rho ^{2}$
which is small since $r\sim 0.04$ and $|\rho|\sim 0.25$.

\section{Lepton and CP asymmetries}

\subsection{Forward-backward asymmetries}

The differential and normalized forward-backward asymmetries (FBAs) for the
decays of $\Lambda _{b}\to \Lambda l^{+}l^{-}$ as a function of $s$ are
defined by
\be
\frac{dA_{FB}\left( s\right) }{ds}=\left[ \int_{0}^{1}d\hat{z}\ \frac{%
d^{2}\Gamma \left( s,\hat{z}\right) }{dsd\hat{z}}-\int_{-1}^{0}d\hat{z}\
\frac{d^{2}\Gamma \left( s,\hat{z}\right) }{dsd\hat{z}}\right]
\ee
and
\be
{\cal A}_{FB}\left( s\right) =\frac{1}{d\Gamma \left( s\right) /ds}\left[
\int_{0}^{1}d\hat{z}\frac{d^{2}\Gamma \left( s,\hat{z}\right) }{dsd\hat{z}}%
-\int_{-1}^{0}d\hat{z}\frac{d^{2}\Gamma \left( s,\hat{z}\right) }{dsd\hat{z}}%
\right] \,,
\ee
respectively. Explicitly, using Eq. (\ref{dorate}), we obtain
\begin{equation}
\frac{dA_{FB}\left( s\right) }{ds}=\frac{G_{F}^{2}\alpha _{em}^{2}\lambda
_{t}^{2}}{2^{8}\pi ^{5}}M_{\Lambda _{b}}^{5}\phi \left( s\right) \left( 1-4%
\frac{\hat{m}_{l}^{2}}{\hat{s}}\right) \bar{f}^{2}R_{FB}\left( s\right)
\label{dfba}
\end{equation}
and
\begin{equation}
{\cal A}_{FB}\left( s\right) =\frac{3}{2}\sqrt{\phi \left( s\right) }\sqrt{1-%
\frac{4\hat{m}_{l}^{2}}{s}}\frac{R_{FB}\left( s\right) }{R_{\Lambda
_{b}}\left( s\right) }  
\label{fba}
\end{equation}
where

\begin{eqnarray}
R_{FB}\left( s\right) &=&s\left[ 1-2\sqrt{r}\rho -\left( 1-r\right) \rho
^{2}\right] {\rm {Re}C_{9}^{eff}C_{10}^{*}}  \nonumber \\
&&+2\hat{m}_{b}\left( 1-s\rho ^{2}\right) {\rm {Re}C_{7}^{eff}C_{10}^{*}}.
\label{rfb}
\end{eqnarray}
It is known that the
 FBA is a parity-odd but CP-even observable, which depends on the
chirality of
the leptonic and hadronic currents. In order to obtain one
power of $%
\hat{z}$ dependence, the related differential decay rate should be
associated with $TrL_{\mu }L_{\nu }^{5}$. This explains why the FBAs depend
on $ReC_{9}^{eff}C_{10}^{*}$ and $ReC_{7}^{eff}C_{10}^{*}$.
However, unlike that in the decays of $B\rightarrow Kl^{+}l^{-}$ where
the FBAs are always zero
since they only involve vector and tensor types of currents, 
the transition matrix elements in the baryonic decays 
preserve the chirality of free quark
interaction.

Similar to the B-meson decays \cite{Bmeson,Burdman}
the FBA in Eq. (\ref{fba})
vanishes at $s_{0}$ which
satisfies with the relation
\be
{\rm {Re}C_{9}^{eff}C_{10}^{*}=-}\frac{2\hat{m}_{b}}{s_{0}}\frac{1-s_{0}\rho
^{2}}{1-2\sqrt{r}\rho -\left( 1-r\right) \rho ^{2}}{\rm {Re}%
C_{7}^{eff}C_{10}^{*}.}
\label{s0}
\ee
We will see later that the vanishing point is only sensitive to the effects
of weak interaction.

\subsection{Lepton polarization asymmetries}

To display the spin effects of the lepton, we choose the four-spin vector
of $l^{+}$ in terms of a unit vector, $\hat{\xi}$, along the
spin of $l^{+}$ in its rest frame, as
\begin{equation}
s_{+}^{0}\,=\,\frac{\vec{p}_{+}\cdot \hat{\xi}}{m_{l}},\qquad \vec{s}%
_{+}\,=\,\hat{\xi}+\frac{s_{+}^{0}\,}{E_{l^{+}}+m_{l}}\vec{p}_{+} \,,
\label{sv}
\end{equation}
and the unit vectors along the longitudinal and transverse components of the
$l^{+}$ polarization to be
\begin{eqnarray}
\hat{e}_{L} &=&\frac{\vec{p}_{+}}{\left| \vec{p}_{+}\right| },  \nonumber \\
\hat{e}_{T} &=&\frac{\vec{p}_{\Lambda }\times \vec{p}_{+}}{\left| \vec{p}%
_{\Lambda }\times \vec{p}_{+}\right| }\,,  \nonumber \\
\hat{e}_{N} &=&\hat{e}_{L}\times \hat{e}_{T}\,,  \label{uv}
\end{eqnarray}
respectively.

Defining the longitudinal and transverse $l^{+}$ polarization asymmetries by
\begin{equation}
P_{i}\left( \hat{s}\right) =\frac{d\Gamma \left( \hat{e}_{i}\cdot \hat{\xi}%
=1\right) -d\Gamma \left( \hat{e}_{i}\cdot \hat{\xi}=-1\right) }{d\Gamma
\left( \hat{e}_{i}\cdot \hat{\xi}=1\right) +d\Gamma \left( \hat{e}_{i}\cdot
\hat{\xi}=-1\right) }\,,  \label{pasy}
\end{equation}
with $i=L$ and $T$, we find that

\begin{eqnarray}
P_{L} &=&\sqrt{1-\frac{4m_{l}^{2}}{q^{2}}}\frac{R_{L}(s)}{R_{\Lambda _{b}}(s)%
}  \label{pl} \,,\\
P_{T} &=&\frac{3}{4}\pi \hat{m}_{l}\sqrt{1-\frac{4m_{l}^{2}}{q^{2}}}\sqrt{%
s\phi \left( s\right) }\frac{R_{T}(s)}{R_{\Lambda _{b}}(s)}\,,  \label{pt}
\end{eqnarray}
where
\begin{eqnarray}
R_{L} &=&-{\rm {Re}C_{9}^{eff}C_{10}^{*}\left[ \left( 1-r\right)
^{2}+s\left( 1+r\right) -2s^{2}+6\sqrt{r}\rho s\left( 1-r+s\right) \right. }
\nonumber \\
&&\left. +\rho ^{2}s\left( 2\left( 1-r\right) ^{2}-s\left( 1+r\right)
-s^{2}\right) \right]   \nonumber \\
&&-6\hat{m}_{b}{\rm {Re}C_{7}^{eff}C_{10}^{*}\left[ \left( 1-r-s\right)
\left( 1+\rho ^{2}s\right) +4\sqrt{r}\rho s\right] },  \label{rl} \\
R_{T} &=&\left[ 1-2\sqrt{r}\rho -\rho ^{2}\left( 1-r\right) \right] {\rm {Im}%
C_{9}^{eff}C_{10}^{*}}  \nonumber \\
&&+\frac{2\hat{m}_{b}}{s}\left( 1-\rho ^{2}s\right) {\rm {Im}%
C_{7}^{eff}C_{10}^{*}.}  \label{rt}
\end{eqnarray}
Here we do not discuss the normal polarization ($P_{N}$) because
the  nonperturbative effects 
from the form factors are large at the
 small $s$ region
and moreover, the dependence of Wilson coefficients is similar to the
invariant mass distribution \cite{chen2}.
We note that the longitudinal lepton polarization of $P_L$ in 
Eq. (\ref{pl}) is also a parity-odd and CP-even observable just like the
FBA, whereas
$P_T$ in Eq. (\ref{pt}) a T-odd one which is related to the triple
correlation of $\vec{s}_+\cdot (\vec{p}_{\Lambda}\times \vec{p}_+)$.
In general, $P_T$ can be induced without CP violation as the cases in
 B-meson \cite{Kao} and kaon \cite{geng-k} decays. However, we expect
that they are small. Moreover, such effects can be extracted away
while we consider the difference between the particle and anti-particle as
discussed in the next section.

\subsection{CP asymmetries}

In this subsection, we define the following interesting 
direct CP asymmetries (CPAs) by
\begin{eqnarray}
\Delta _{\Gamma } &=&\frac{d\Gamma -d\bar{\Gamma}}{d\Gamma
+d\bar{\Gamma}}\,,
\label{rate1cp} \\
\Delta _{FB} &=&\frac{d\Gamma _{FB}-d\bar{\Gamma}_{FB}}{d\Gamma +d\bar{\Gamma%
}}\,,  \label{fb1cp} \\
\Delta _{P_{i}} &=&\frac{d\Gamma \left( \vec{\xi}\cdot \vec{e}_{i}\right) -d%
\bar{\Gamma}\left( \vec{\xi}\cdot \vec{e}_{i}\right) }{d\Gamma +d\bar{\Gamma}%
},{\rm {\quad }i=L,T}  \label{pltcp}
\end{eqnarray}
where we have used $d\Gamma +d\bar{\Gamma}$ as the normalization.
The above four CPAs are CP-odd quantities and they are CP violating
observables.
 For $\Delta _{\Gamma,FB,P_L}$ in Eqs. (\ref{rate1cp})-(\ref{pltcp}),
 to display the difference of
the physical observable between the particle and anti-particle, it is
necessary to have the strong and weak phases simultaneously in the
processes. In the decays of $b\rightarrow sl^{+}l^{-}\ (l=e,\mu,\tau)$,
the strong phases are generated by the absorptive parts of one-loop matrix
elements in operators $O_{1}\sim O_{6}$ and
LD contributions. However, since $P_{T}$ is a
$T$-odd observable and only related to the imaginary couplings, even
without strong phases, we still can have nonzero values of
$\Delta_T ( \Lambda_b\rightarrow \Lambda l^{+}l^{-}) $.
For $b\rightarrow sl^{+}l^{-}$
and $\bar{b}\rightarrow \bar{s}l^{-}l^{+}$ decays
the Wilson coefficients $C_{9}^{eff}(\mu )$ and $C_{7}^{eff}(\mu )$
in Eqs. (\ref{ceff9}) and (\ref{ceff7})
can be rewritten as
\begin{eqnarray}
C_9^{eff}(\mu ) &=&C_9^0(\mu )+iC_9^{abs}(\mu )\,,
\nn \\
\bar{C}_9^{eff}(\mu ) &=&C_9^{0*}(\mu )+iC_9^{abs}(\mu )\,,
\nn \\
C_7^{eff}(\mu ) &=&C_7^0(\mu )+iC_7^{abs}(\mu ) \,,
\nn \\
\bar{C}_7^{eff}(\mu ) &=&C_7^{0*}(\mu )+iC_7^{abs}(\mu )\,,
\end{eqnarray}
with
\begin{eqnarray}
C_9^0(\mu ) &=&C_9(\mu )+{\rm {Re}}Y\left( z,s^{\prime }\right)\,,
\nn\\
C_{7}^{0}(\mu ) &=&C_{7}(\mu )+{\rm {Re}}C_{7}^{\prime }(\mu ,q^{2})\,,
\nn \\
C_{9}^{abs}(\mu ) &=&{\rm {Im}}Y\left( z,s^{\prime }\right)\,,
\nn \\
C_{7}^{abs}(\mu ) &=&{\rm {Im}}C_{7}^{\prime }(\mu ,q^{2})\,,
\end{eqnarray}
where we have assumed that the strong phases are all from the SM
and there are no weak phases in absorptive parts.
We note that there is no strong phase in $C_{10}$.

According to Eqs. (\ref{rate}), (\ref{fba}), (\ref{pl}), and (\ref{pt}),
the CP asymmetries are all related to the following combinations:

\begin{eqnarray}
 ReC_9^{eff}C_7^{eff*}-Re\bar{C}_9^{eff}\bar{C}_7^{eff*}
&=&2C_9^{abs}ImC_7^0+2C_7^{abs}ImC_9^0\,,
\nn \\
ReC_{7,9}^{eff}C_{10}^*-Re\bar{C}_{7,9}^{eff}\bar{C}_{10}^*
&=&2C_{7,9}^{abs}ImC_{10}\,,
\nn \\
ImC_{7,9}^{eff}C_{10}^*-Im_{7,9}^{eff}\bar{C}_{10}^{*} &=&2
ImC_{7,9}^{0}C_{10}^{*}+2C_{7,9}^{abs}{Im}C_{10}\,,
\nn \\
\left| C_{7,9}^{eff}\right| ^{2}-\left| \bar{C}_{7,9}^{eff}\right| ^{2}
&=&4C_{7,9}^{abs}{Im}C_{7,9}^{0}\,.
\end{eqnarray}
Explicitly,
the CP asymmetries in Eqs. (\ref{rate1cp}), (\ref{fb1cp}), and
(\ref{pltcp})
are found to be
\begin{eqnarray}
\Delta _{\Gamma } &=&\frac{2}{R_{\Lambda _{b}}\left( s\right) }\left\{ 6\hat{%
m}_{b}\left( 1+2\frac{m_{l}^{2}}{q^{2}}\right) \left[ 2\sqrt{r}\rho s+\left(
t-r\right) \left( 1+s\rho ^{2}\right) \right] \left[ C_{9}^{abs}{\rm {Im}%
C_{7}^{0}+C_{7}^{abs}{Im}C_{9}^{0}}\right] \right.
 \nonumber \\
&&+\left[ -2r\left( 1+2\frac{m_{l}^{2}}{q^{2}}\right) -4t^{2}\left( 1-\frac{%
m_{l}^{2}}{q^{2}}\right) +3\left( 1+r\right) t+6\hat{m}_{l}^{2}t\right]
\nn\\
&& \times \left[ 4\hat{m}_{b}^{2}\rho ^{2}C_{7}^{abs}{\rm
{Im}C_{7}^{0}+C_{9}^{abs}{Im}%
C_{9}^{0}}\right]
 \nonumber \\
&&+\left[ -2t^{2}\left( 1+2\frac{m_{l}^{2}}{q^{2}}\right) -4r\left( 1-\frac{%
m_{l}^{2}}{q^{2}}\right) +3\left( 1+r\right) t-6\frac{\hat{m}_{l}^{2}}{s}%
\left( 2r-t-tr\right) \right]
\nonumber \\
&&\times \left[ \frac{4\hat{m}_{b}^{2}}{s}C_{7}^{abs}{\rm {Im}%
C_{7}^{0}+s\rho ^{2}C_{9}^{abs}{Im}C_{9}^{0}}\right] +6\sqrt{r}\rho \left(
1-t\right) \left( 1+2\frac{m_{l}^{2}}{q^{2}}\right)  \nonumber \\
&&\left. \times \left[ 4\hat{m}_{b}^{2}C_{7}^{abs}{Im}
C_{7}^{0}+sC_9^{abs}{Im}C_9^0\right] -6\sqrt{r}s\delta \left( 1+2%
\frac{m_l^2}{q^2}\right) C_9^{abs} {Im}C_9^0\right\} ,
\label{ratecp} \\
\Delta_{FB} &=&\frac{3}{2R_{\Lambda_b}(s)}
\sqrt{1-\frac{4m_l^2}{q^2}}
\phi (s)ImC_{10}\left[ s\left( 1-2\sqrt{r}\rho -\left( 1-r\right)
\rho^2\right) C_9^{abs}\right.
\nn\\
&&+
\left. 2\hat{m}_{b}\left( 1-s\rho ^{2}\right)
C_{7}^{abs}\right]  \,,
\label{fbacp} \\
\Delta_{P_L} &=&-\frac{1}{R_{\Lambda _{b}}(s)}\sqrt{1-\frac{4m_{l}^{2}}{%
q^{2}}} {Im}C_{10}\left\{ C_{9}^{abs}\left[ 6\sqrt{r}\rho s\left(
1-r+s\right) +\left( 1+2s\rho ^{2}\right) \left( 1-r\right) ^{2}\right.
\right.   \nonumber \\
&& \left. +s\left( 1-s\rho ^{2}\right) \left( 1+r\right) -s^{2}\left(
2+s\rho ^{2}\right) \right]
\nn\\
&&
\left.+6\hat{m}_bC_7^{abs}\left[ \left(
1-r-s\right) \left( 1+s\rho ^{2}\right) +4\sqrt{r}\rho s\right]
\right\}\,,
\label{plcp} \\
\Delta _{P_{T}} &=&\frac{3\pi \hat{m}_{l}}{4R_{\Lambda _{b}}(s)}\sqrt{1-%
\frac{4m_{l}^{2}}{q^{2}}}\sqrt{s\phi \left( s\right) }\left\{ \left( {Im}%
C_{9}^{0}C_{10}^{*}+C_{9}^{abs}{Im}C_{10}\right) \right.
\nn\\
&&\times\left( 1-2\sqrt{r}\rho
-\left( s+2t-2r\right) \rho ^{2}\right)
\nonumber \\
&&\left.
+\frac{2\hat{m}_b}{s}\left( 1-s\rho ^2\right)
\left( {Im}
C_7^0C_{10}^*+C_7^{abs}{Im}C_{10}\right) \right\}\,.
\label{ptcp}
\end{eqnarray}
As seen from the above equations,
$\Delta _{\Gamma }$ is related to ${\rm {Im}C_{7}}$ and ${\rm {Im}C_{9}}$,
while $\Delta _{FB}$, $\Delta _{P_{L}}$ and $\Delta_{P_T}$
depend on $ImC_{10}$.
Moreover, for small values of $C_{9}^{abs}{Im}C_{10}$ and
$C_{7}^{abs}{Im}C_{10}$,
$\Delta_{P_T}$ would still be sizable because
${Im}C_{9}^{0}C_{10}^{*}$ or ${Im}C_{7}^{0}C_{10}^{*}$ would be
large.

\section{Numerical analysis}

In our numerical calculations, the Wilson coefficients are evaluated at the
scale $\mu \simeq m_{b}$ and the other parameters are listed in Table 1 of
Ref. \cite{chen1}. From Eq. (\ref{appff}), we know that
the main effects to the
deviation of the HQET are from $\delta $.
By using a proper nonzero value of $\delta$, we
will see later that the deviations of the decay branching ratios of
$\Lambda
_{b}\rightarrow \Lambda l^{+}l^{-}$  are only a few percent.
Since there is no complete calculation for the form factors of
the $\Lambda
_{b}\rightarrow \Lambda $ transition in the literature, we use the form
factors derived
from QCD sum rule under the assumption of the HQET, given by
\begin{equation}
F_{i}(q^{2})=\frac{F_{i}(0)}{1+aq^{2}+bq^{4}}\,,
\label{qcdff}
\end{equation}
with the parameters shown in Table 1 of Ref. \cite{chen2}.
In order to illustrate the contributions of new physics, we adopt the
results of the generic supersymmetric extension of the SM
\cite{Masiero} in which
\begin{eqnarray}
C_7^{SUSY} &=&-1.75\left( \delta _{23}^u\right)_{LL}-0.25\left( \delta
_{23}^u\right)_{LR}-10.3\left( \delta_{23}^d\right)_{LR},
\nonumber
\\
C_{9}^{SUSY} &=&0.82\left( \delta _{23}^{u}\right) _{LR},  \nonumber \\
C_{10}^{SUSY} &=&-9.37\left( \delta _{23}^{u}\right) _{LR}+1.4\left( \delta
_{23}^{u}\right) _{LR}\left( \delta _{33}^{u}\right) _{RL}+2.7\left( \delta
_{23}^{u}\right) _{LL},
\end{eqnarray}
and take the following values instead of scanning the whole allowed
parameter space:
\be
\left( \delta_{23}^u\right)_{LL} &\sim  & 0.1\,,
\nn \\
\left( \delta _{33}^{u}\right) _{RL} & \sim  & 0.65\,,
\nn \\
\left( \delta _{23}^{d}\right) _{LR} & \sim  & 3\times 10^{-2}e^{i\frac{%
2\pi }{5}}\,,
\nn \\
\left( \delta _{23}^{u}\right) _{LR} & \sim  & -0.8e^{i\frac{\pi }{4}}\,,
\label{delta}
\ee
where $(\delta_{ij}^{q})_{AB}$ $(i,j=1,2,3$ and $A,B=L,R)$
denote the parameters in the mass insertion method, which
describe the effects of the flavour violation. The
set of the parameters in Eq. (\ref{delta}) satisfies with the
constraint from $B\rightarrow
X_{s}\gamma $ on $C_{7}=C_{7}^{SM}+C_{7}^{SUSY}$ \cite{Masiero}. Hence, the
numerical values of the SUSY contributions to the relevant Wilson
coefficients are as follows:
\be
Re\,C_7^{SUSY}\;\simeq\;  0.06\,, && Im\,C_{7}^{SUSY}\;\simeq\; -0.29\,,
\nn\\
Re\,C_9^{SUSY} \;\simeq \;-0.46\,, && Im\,C_{9}^{SUSY}\;\simeq\; -0.46\,,
\nn \\
Re\,C_{10}^{SUSY}\; \simeq\; 4.78\,,&& Im\,C_{10}^{SUSY}
\;\simeq\; 4.50\,.
\label{npv}
\ee
We note that the contributions of the minimal supersymmetric standard
model (MSSM) to $b\rightarrow sl^{+}l^{-}$ can be found in
Refs. \cite{BBMR} and \cite{AGM}.

To show the typical values of various asymmetries, we
define the
integrated quantities as
\begin{equation}
\bar{Q}=\int_{s_{\min }}^{s_{\max }}Q(s)ds  \label{ave}
\end{equation}
where $Q$ denote the physical observables with $s_{\min
}=4\hat{m}_{l}$ and $s_{\max }=(1-\sqrt{r})^{2}$.

\subsection{Decay rates and invariant mass distributions}

We now discuss the influences of $\delta $,
$\rho $, and $\omega $ on the branching ratios (BRs) of $\Lambda
_{b}\rightarrow \Lambda l^{+}l^{-}$ decays in detail. The effects of
$k_{j}$ for compensating the assumption of the FA and VMD
 have been analyzed in \cite{chen1}.
In Table \ref{brpa}, we show the BRs by choosing different sets of
parameters.
Our results are given as follows:

\begin{table}[h]
\caption{BRs (in the unit of $10^{-6}$) for various parameters with
$\omega=0$ and neglecting LD effects.}
\label{brpa}
\begin{center}
\begin{tabular}{|l|l|l|l|}
\hline
Parameter & $\Lambda _{b}\to \Lambda e^{+}e^{-}$ & $\Lambda _{b}\to \Lambda
\mu ^{+}\mu ^{-}$ & $\Lambda _{b}\to \Lambda \tau ^{+}\tau ^{-}$ \\
\hline\hline
$HQET$ & \multicolumn{1}{|c|}{$2.23$} & \multicolumn{1}{|c|}{$2.08$} &
\multicolumn{1}{|c|}{$1.79\times 10^{-1}$} \\ \hline
$\delta =0.05$ & \multicolumn{1}{|c|}{$2.36$} & \multicolumn{1}{|c|}{$2.21$}
& \multicolumn{1}{|c|}{$1.86\times 10^{-1}$} \\ \hline
$\delta =-0.05$ & \multicolumn{1}{|c|}{$2.09$} & \multicolumn{1}{|c|}{$1.96$}
& \multicolumn{1}{|c|}{$1.71\times 10^{-1}$} \\ \hline
$\rho =0,\delta =0$ & \multicolumn{1}{|c|}{$2.52$} & \multicolumn{1}{|c|}{$%
2.38$} & \multicolumn{1}{|c|}{$2.66\times 10^{-1}$} \\ \hline
$C_{7}=0,\delta =0$ & \multicolumn{1}{|c|}{$2.36$} & \multicolumn{1}{|c|}{$%
2.34$} & \multicolumn{1}{|c|}{$2.23\times 10^{-1}$} \\ \hline
$C_{7}=-C_{7}^{SM},\delta =0$ & \multicolumn{1}{|c|}{$3.34$} &
\multicolumn{1}{|c|}{$3.19$} & \multicolumn{1}{|c|}{$2.76\times 10^{-1}$} \\
\hline
\end{tabular}
\end{center}
\end{table}

\begin{enumerate}
\item  By taking $\left| \delta \right| =0.05$ which means $10\%$ 
away from that in to the HQET, we clearly see that the deviations of the
BRs are only $4-6\%$. It is a good approximation to neglect the explicit
$\delta $ term in Eqs. (\ref{I0}), (\ref{rate1}) and (\ref{ratecp}).
Hence, $\bar{f}=\left(f_{1}+g_{1}\right) /2$, which also owns the $\delta
$ effect, is the main nonperturbative part.

\item  If $\rho =0$, the effects are about $10\%$ for $e$ and $\mu $
modes but $48\%$ for $\tau $ one.

\item  If one neglects the contribution from $C_{7}$, the
influences on $B(\Lambda_{b}\to \Lambda l^{+}l^{-})$ for $e$, $\mu$ and
$\tau$ modes are about $5\%$, $12\%$ and $24\%$, respectively. However,
taking the magnitude of $C_{7}$ is the same as the SM but with an opposite
sign, the deviations are all over $50\%$.
\end{enumerate}

The contributions of the parameter $\omega $ to the BRs of
$\Lambda _{b}\to \Lambda l^{+}l^{-}$  are listed in Table \ref{bromega}
and the invariant mass distributions are shown in Figure 1. From
Table \ref{bromega}, it is clear that the nonfactorizable effects are small
on the BRs. However, those directly related to $\omega $ effects
such as $P_{T}$ and CP asymmetries will have large influences.
\begin{table}[h]
\caption{BRs (in the unit of $10^{-6}$) without LD effects with different
values of $\omega $%
}
\label{bromega}
\begin{center}
\begin{tabular}{|l|l|l|l|}
\hline
Mode & $\Lambda _{b}\to \Lambda e^{+}e^{-}$ & $\Lambda _{b}\to \Lambda \mu
^{+}\mu ^{-}$ & $\Lambda _{b}\to \Lambda \tau ^{+}\tau ^{-}$ \\ \hline\hline
$\omega =0.15$ & \multicolumn{1}{|c|}{$2.24$} & \multicolumn{1}{|c|}{$2.12$}
& \multicolumn{1}{|c|}{$1.89\times 10^{-1}$} \\ \hline
$\omega =0.$ & \multicolumn{1}{|c|}{$2.23$} & \multicolumn{1}{|c|}{$2.08$} &
\multicolumn{1}{|c|}{$1.79\times 10^{-1}$} \\ \hline
$\omega =-0.15$ & \multicolumn{1}{|c|}{$2.25$} & \multicolumn{1}{|c|}{$2.06$}
& \multicolumn{1}{|c|}{$1.71\times 10^{-1}$} \\ \hline
\end{tabular}
\end{center}
\end{table}

As for the new physics contributions, using the values of the SUSY model 
in Eq. (\ref{npv}), we show the results in Table \ref{newph}. Although the
deviations of the BRs to the SM are not significant, they have a large
effect on the lepton and CP asymmetries which will be shown next.

\begin{table}[h]
\caption{BRs (in unit of $10^{-6}$) in the generic SUSY model.}
\label{newph}
\begin{center}
\begin{tabular}{|l|l|l|l|}
\hline
Model & $\Lambda _{b}\to \Lambda e^{+}e^{-}$ & $\Lambda _{b}\to \Lambda \mu
^{+}\mu ^{-}$ & $\Lambda _{b}\to \Lambda \tau ^{+}\tau ^{-}$ \\ \hline\hline
SUSY & \multicolumn{1}{|c|}{$2.47$} & \multicolumn{1}{|c|}{$2.24$} &
\multicolumn{1}{|c|}{$1.79\times 10^{-1}$} \\ \hline
\end{tabular}
\end{center}
\end{table}

\subsection{Forward-backward and lepton polarization asymmetries}

 From Eq. (\ref{HQETff}) and $R=F_{2}/F_{1}\simeq-0.25$
in the HQET, we have that $\rho \simeq
-0.26$.
We note that $\rho $ is defined by the ratio of the form factors
and it is expected to be
insensitive to the QCD models. With $s_{\max }\simeq 0.64$, we obtain $%
s_{\max }\rho ^{2}\simeq 0.04$, $(1-r)\rho ^{2}\simeq 0.06$ and
$2\sqrt{r}\rho $ $\simeq 0.2$.
 Using these values, one can simplify  Eqs. (\ref{rfb}) and (\ref{s0})
to
\begin{equation}
R_{FB}\left( s\right) \simeq s\left( 1-2\sqrt{r}\rho \right) {\rm {Re}%
C_{9}^{eff}C_{10}^{*}}+2\hat{m}_{b}{\rm {Re}C_{7}^{eff}C_{10}^{*}}
\label{refb}
\end{equation}
and
\begin{equation}
{\rm {Re}C_{9}^{eff}C_{10}^{*}\simeq -}\frac{2\hat{m}_{b}}{s_{0}\left(
1-2%
\sqrt{r}\rho \right) }{\rm {Re}C_{7}^{eff}C_{10}^{*}}\,,  \label{va}
\end{equation}
respectively.
It is easy to see that
$s_{0}$ is only sensitive to the Wilson coefficients. The result
is similar to the case in $B\rightarrow K^{*}l^{+}l^{-}$ decay
\cite{Bmeson,Burdman} where the approximation of the large energy
effective
theory (LEET) \cite{Charles} is used. As for the lepton polarization
asymmetries,
with the same approximation, Eqs. (\ref{rl}) and (\ref{rt}) can also be
reduced to
\begin{eqnarray}
R_{L} &\simeq &-{\rm {Re}C_{9}^{eff}C_{10}^{*}\left[ 1+s-2s^{2}+6\sqrt{r}%
\rho s\left( 1+s\right) \right] }  \nonumber \\
&&-6\hat{m}_{b}{\rm {Re}C_{7}^{eff}C_{10}^{*}\left[ 1-s+4\sqrt{r}\rho
s\right] },  \label{repl} \\
R_{T} &\simeq &\left( 1-2\sqrt{r}\rho \right) {\rm
{Im}C_{9}^{eff}C_{10}^{*}%
}+\frac{2\hat{m}_{b}}{s}{\rm {Im}C_{7}^{eff}C_{10}^{*}}\,,
\label{rept}
\end{eqnarray}
respectively.
Hence, the lepton asymmetries are all more sensitive to the Wilson
coefficients than the nonperturbative QCD effects.

It is worth to mention that the effects of $\omega $, introduced for the LD
contributions to $b\rightarrow s\gamma $ and absorbed to $C_{7}^{eff}$, will
change ${\rm {Re}C_{7}^{eff}}$ in the SM such that $s_{0}$ is also
shifted. Therefore, in terms of $s_{0},$ we can also theoretically
determine $\omega$ by comparing the result with that of $\omega =0$.
Another interesting quantity is T-odd observable of $P_{T}$ which is
proportional to $C_{10}{\rm {Im}C_{7}^{eff}}$ in the SM.
Due to the enhancement of $C_{10}$, a nonzero value of
$\omega $ will modify $P_{T}$ enormously. As for the other asymmetries,
the effects are insignificant. The estimations of integrated lepton
asymmetries with different values of $\omega $ in the SM are displayed in
Table \ref{asy} and the corresponding distributions are shown in Figures
$2-4$.

\begin{table}[h]
\caption{Integrated lepton asymmetries in the SM without LD effects.}
\label{asy}
\begin{center}
\begin{tabular}{|l|l|l|l|l|}
\hline
 &  &  &  &  \\
Parameter & Mode & $10^{2}\bar{A}_{FB}$ & $10^{2}\bar{P}_{L}$ &
$10^{2}\bar{P}_{T}$ \\ \hline\hline
$\omega =0.15$ & \multicolumn{1}{|c|}{$\Lambda _{b}\to \Lambda \mu ^{+}\mu
^{-}$} & \multicolumn{1}{|r|}{$-14.37$} & \multicolumn{1}{|r|}{$59.50$} &
\multicolumn{1}{|r|}{$0.11$} \\ \cline{2-5}
& $\Lambda _{b}\to \Lambda \tau ^{+}\tau ^{-}$ & \multicolumn{1}{|r|}{$-3.98$%
} & \multicolumn{1}{|r|}{$10.70$} & \multicolumn{1}{|r|}{$0.53$} \\ \hline
$\omega =0.$ & \multicolumn{1}{|c|}{$\Lambda _{b}\to \Lambda \mu ^{+}\mu
^{-} $} & \multicolumn{1}{|r|}{$-13.38$} & \multicolumn{1}{|r|}{$58.30$} &
\multicolumn{1}{|r|}{$0.07$} \\ \cline{2-5}
& $\Lambda _{b}\to \Lambda \tau ^{+}\tau ^{-}$ & \multicolumn{1}{|r|}{$-3.99$%
} & \multicolumn{1}{|r|}{$10.84$} & \multicolumn{1}{|r|}{$0.39$} \\ \hline
$\omega =-0.15$ & \multicolumn{1}{|c|}{$\Lambda _{b}\to \Lambda \mu ^{+}\mu
^{-}$} & \multicolumn{1}{|r|}{$-12.24$} & \multicolumn{1}{|r|}{$56.70$} &
\multicolumn{1}{|r|}{$0.04$} \\ \cline{2-5}
& $\Lambda _{b}\to \Lambda \tau ^{+}\tau ^{-}$ &
\multicolumn{1}{|r|}{$-4.00$}  & \multicolumn{1}{|r|}{$10.94$} &
\multicolumn{1}{|r|}{$0.23$} \\ \hline
\end{tabular}
\end{center}
\end{table}

To illustrate the new physics effects, the integrated lepton asymmetries
in the generic SUSY model with $\omega$ 
are listed in Table \ref{susyasy} and their distributions 
as a function of $q^2/M_{\Lambda_b}$ are shown
 in Figures $5-7$. From the figures, we see that SUSY effects make
the shapes of lepton asymmetries quite differ from that in the SM. We
summary the results as follows:

\begin{enumerate}
\item Since the SD contributions to $C_{9}C_{10}^{*}$ and
$ReC_{7}C_{10}^{*}$ are $-1.40$ and $-1.35$, respectively,
 which violate the condition in Eq. (\ref{va}),
the vanishing point is removed.

\item  Due to the factor of $\hat{m}_{b}/s$, from Figure $7$,
we see that $ImC_{7}^{eff}C_{10}^{*}$ has a large effect on $P_{T}$
in the small $s$ region.

\item  In the SUSY model, $P_{T}$ could reach $1\%$ and
$10\%$ for the light lepton and $\tau $ modes, which  are only
$0.2\%$ and $3\%$ at most in the SM, respectively.
\end{enumerate}

\begin{table}[h]
\caption{Integrated lepton asymmetries in the generic SUSY model
with $\omega=0$.}
\label{susyasy}
\begin{center}
\begin{tabular}{|l|l|l|l|}
\hline
&  &  &  \\
Mode & $10^{2}\bar{A}_{FB}$ & $10^{2}\bar{P}_{L}$ & $10^{2}\bar{P}_{T}$ \\
\hline
\hline
\multicolumn{1}{|c|}{$\Lambda _{b}\to \Lambda \mu ^{+}\mu ^{-}$} &
\multicolumn{1}{|r|}{$-10.53$} & \multicolumn{1}{|r|}{$24.46$} &
\multicolumn{1}{|r|}{$-0.57$} \\ \cline{1-4}
$\Lambda _{b}\to \Lambda \tau ^{+}\tau ^{-}$ & \multicolumn{1}{|r|}{$-1.84$}
& \multicolumn{1}{|r|}{$4.40$} & \multicolumn{1}{|r|}{$-2.51$} \\ \hline
\end{tabular}
\end{center}
\end{table}

\subsection{CP asymmetries}

In the SM,
for $b\rightarrow sl^{+}l^{-}$, the relevant CKM matrix element is
$V_{tb}V_{ts}^{*}$ which is real under the Wolfenstein's parametrization.
Nonzero CPAs will indicate clearly the existence of new
physics. 
We remark that
the CPAs can be in fact induced by the complex CKM matrix element
$V_{ub}V_{us}^{*}$  which is also the source of the direct CPA in
$B\rightarrow X_{s}\gamma $ in the SM.
 However, we expect that such effects to the CPAs in
$b\rightarrow sl^{+}l^{-}$ are smaller than that 
in $B\rightarrow X_{s}\gamma $ where the CPA is less than $1\%$.
The main reason for the smallness is because
of the presences of $C_9$ and $C_{10}$
contributions to the rates of
$b\rightarrow sl^{+}l^{-}$, which are absent in $B\to X_{s}\gamma$.

With the values in Eq. (\ref{npv}), the averaged CPAs in the generic SUSY
model for $\Lambda _{b}\to \Lambda l^{+}l^{-}$ are listed in Table
\ref{cpasy} and their distributions as a function of
$s=q^2/M_{\Lambda_b}^2$ are shown
in Figures $8-11$. The results are given as follows:

\begin{enumerate}
\item  From Eqs. (\ref{ratecp}) and (\ref{ptcp}),
we see that the terms corresponding to $C_{7}^{abs}ImC_{7}^{0}$ and
$ImC_{7}^{0}C_{10}^{*}+C_{7}^{abs}{Im}C_{10}$
 are associated with a factor of $\hat{m}_{b}/s$. If
sizable imaginary parts exist, in the small $s$ region the distributions
will be significant. Due to this reason, 
in Figure 8a one finds that $\Delta _{\Gamma }(s)$ for
$\Lambda_{b}\to \Lambda l^{+}l^{-}$ $(l=e,\mu )$
 increase as $s$ decreases. On the other hand, if the term with
 $\hat{m}_{b}/s$ in Eq. (\ref{ptcp}) is dropped, the distributions of
$\Delta_{P_{T}}(s)$ for $e$ and $\mu$ modes do not contain zero
value. We note that with the values in Eq. (\ref{npv}), the main effect on
$\Delta_{\Gamma }(s)$ in the small $s$ region is from 
$C_{b\rightarrow s\gamma}^{\prime }$.

\item  $\Delta_{P_L}(s)$ for all lepton channels and $\Delta
_{P_{T}}(s)$ for the $\tau $ one could be over $10\%$,
while the remaining CP asymmetries are at the level of a few percent.
We remark that if we can scan all the allowed SUSY parameters, the
asymmetries except $\Delta_{P_{T}}(s)$ for lighter lepton modes
 would  reach up $10\%$.

\item  It is known that $\Delta _{P_{T}}(s)$ is a T-odd
observable and the other CPAs belong to the direct CP violation which
needs absorptive parts in the processes. This is the reason why the
distributions of $\Delta _{\Gamma }(s)$, $\Delta_{FBA}(s)$ and 
$\Delta_{P_{L}}(s)$ around the $RE$ region have the
similar shapes but are different from that of $\Delta _{P_{T}}(s)$. 
Moreover, all the direct CPAs are sensitive to $\omega $ unlike the cases
of the CP conserving lepton asymmetries discussed in Sec. 5.2.
\end{enumerate}

\begin{table}[tbp]
\caption{CP asymmetries in the generic SUSY model for different
values of $\omega$}
\label{cpasy}
\begin{center}
\begin{tabular}{|l|l|l|l|l|l|}
\hline
 & & & & \\
Parameter & Mode & $10^{2}\bar{\Delta}_{\Gamma }$ &
$10^{2}\bar{\Delta}_{FBA}$ &
$10^{2}\bar{\Delta}_{P_{L}}$ & $10^{2}\bar{\Delta}_{P_{T}}$ \\ \hline\hline
$\omega =0.15$ & \multicolumn{1}{|c|}{$\Lambda _{b}\to \Lambda \mu ^{+}\mu
^{-}$} & $2.05$ & \multicolumn{1}{|c|}{$-2.62$} & \multicolumn{1}{|c|}{$6.48$%
} & $-0.53$ \\ \cline{2-6}
& $\Lambda _{b}\to \Lambda \tau ^{+}\tau ^{-}$ & $1.83$ &
\multicolumn{1}{|c|}{$-0.79$} & \multicolumn{1}{|c|}{$1.94$} & $-2.01$ \\
\hline
$\omega =0.$ & \multicolumn{1}{|c|}{$\Lambda _{b}\to \Lambda \mu ^{+}\mu
^{-} $} & $1.59$ & \multicolumn{1}{|c|}{$-1.89$} & \multicolumn{1}{|c|}{$%
5.00 $} & $-0.47$ \\ \cline{2-6}
& $\Lambda _{b}\to \Lambda \tau ^{+}\tau ^{-}$ & $1.38$ &
\multicolumn{1}{|c|}{$-0.59$} & \multicolumn{1}{|c|}{$1.53$} & $-2.21$ \\
\hline
$\omega =-0.15$ & \multicolumn{1}{|c|}{$\Lambda _{b}\to \Lambda \mu ^{+}\mu
^{-}$} & $1.05$ & \multicolumn{1}{|c|}{$-1.06$} & \multicolumn{1}{|c|}{$3.34$%
} & $-0.40$ \\ \cline{2-6}
& $\Lambda _{b}\to \Lambda \tau ^{+}\tau ^{-}$ & $0.89$ &
\multicolumn{1}{|c|}{$-0.37$} & \multicolumn{1}{|c|}{$1.06$} & $-2.41$ \\
\hline
\end{tabular}
\end{center}
\end{table}

\section{Conclusions}

We have given a systematic study on the rare baryonic decays of
$\Lambda_b\to \Lambda l^+l^-\ (l=e,\mu,\tau)$.
For the $\Lambda_b\to \Lambda$ transition, we have related all the form
factors with $F_1$ and  $F_2$, and we have found that 
$\delta=0$ and $\rho\simeq R\equiv F_2/F_1$, in the
limit of the HQET.
Inspired by the HQEF,
we have presented the differential decay rates and the di-lepton
forward-backward, lepton polarization and four possible CP violating
asymmetries in terms of the parameters $\bar{f}$, $\delta$ and $\rho$.
We have shown that the non-factorizable effects for the BRs and CP-even
lepton asymmetries are small but large for $P_T$ and the direct CPAs.
We have also demonstrated that most of the observables such as ${\cal
A}_{FB}$, $P_{L,T}$ and $\Delta_\alpha\ (\alpha=\Gamma, FB,P_L$ and$P_T$).
are insensitive to the non-perturbative QCD effects.
We have illustrated our results in the specific CP violating SUSY model.
We have found that all the direct CP violating asymmetries are in the
level of $1-10\%$. To measure these asymmetries
at the $n\sigma$ level, 
for example, in the tau mode, 
at least $0.5n^2\times (10^9- 10^{10})$ $\Lambda_b$ decays are required.
It could be done in the second generation B-physics experiments,
such as LHCb, ATLAS, and CMS at the LHC, and BTeV at the Tevatron, 
which produce $\sim 10^{12}b\bar{b}$ pairs per year \cite{BB}
Finally we remark that
measuring these CPAs at a level of $10^{-2}$ is a clear indication
of new CP violation mechanism beyond the SM.\\

\noindent
{\bf Acknowledgments}

This work was supported in part by 
the National Science Council of the Republic of China
under Contract Nos. NSC-89-2112-M-007-054 and
NSC-89-2112-M-006-033 and the National Center for Theoretical Science.

\newpage

\appendix{}

\begin{center}
{\Large \bf Appendix}
\end{center}

\noindent
{\large
\bf A. Form factors and decay amplitudes}\\

For the exclusive decays involving $\Lambda _{b}(p_{\Lambda
_{b}})\rightarrow \Lambda ( p_{\Lambda }) $, the transition
form factors can be parametrized generally as follows:

\begin{eqnarray}
\left\langle \Lambda \right| \bar{s}\ \gamma _{\mu }\ b\left| \Lambda
_{b}\right\rangle &=&f_{1}\bar{u}_{\Lambda }\gamma _{\mu }u_{\Lambda
_{b}}+f_{2}\bar{u}_{\Lambda }i\sigma _{\mu \nu }\ q^{\nu }u_{\Lambda
_{b}}+f_{3}q_{\mu }\bar{u}_{\Lambda }u_{\Lambda _{b}},  \label{vc} \\
\left\langle \Lambda \right| \bar{s}\ \gamma _{\mu }\gamma _{5}\ b\left|
\Lambda _{b}\right\rangle &=&g_{1}\bar{u}_{\Lambda }\gamma _{\mu }\gamma
_{5}u_{\Lambda _{b}}+g_{2}\bar{u}_{\Lambda }i\sigma _{\mu \nu }\ q^{\nu
}\gamma _{5}u_{\Lambda _{b}}+g_{3}q_{\mu }\bar{u}_{\Lambda }\gamma
_{5}u_{\Lambda _{b}},  \label{ac} \\
\left\langle \Lambda \right| \bar{s}i\sigma _{\mu \nu }b\left| \Lambda
_{b}\right\rangle &=&f_{T}\bar{u}_{\Lambda }i\sigma _{\mu \nu }u_{\Lambda
_{b}}+f_{T}^{V}\bar{u}_{\Lambda }\left( \gamma _{\mu }q_{\nu }-\gamma _{\nu
}q_{\mu }\right) u_{\Lambda _{b}}  \nonumber \\
&&+f_{T}^{S}\left( P_{\mu }q_{\nu }-P_{\nu }q_{\mu }\right) \bar{u}_{\Lambda
}u_{\Lambda _{b}},  \label{tc} \\
\left\langle \Lambda \right| \bar{s}i\sigma _{\mu \nu }\gamma _{5}b\left|
\Lambda _{b}\right\rangle &=&g_{T}\bar{u}_{\Lambda }i\sigma _{\mu \nu
}\gamma _{5}u_{\Lambda _{b}}+g_{T}^{V}\bar{u}_{\Lambda }\left( \gamma _{\mu
}q_{\nu }-\gamma _{\nu }q_{\mu }\right) \gamma _{5}u_{\Lambda _{b}}
\nonumber \\
&&+g_{T}^{S}\left( P_{\mu }q_{\nu }-P_{\nu }q_{\mu }\right) \bar{u}_{\Lambda
}\gamma _{5}u_{\Lambda _{b}},  \label{atc}
\end{eqnarray}
where $P=p_{\Lambda _{b}}+p_{\Lambda }$ , $q=p_{\Lambda
_{b}}-p_{\Lambda }$ and form factors, $\left\{ f_{i}\right\} $ and
$\{g_{i}\}$, are all functions of $q^{2}$. Using the equations of
the motion, we have
\begin{eqnarray}
\left( M_{\Lambda }+M_{\Lambda _{b}}\right) \bar{u}_{\Lambda }\gamma _{\mu
}u_{\Lambda _{b}} &=&\left( p_{\Lambda _{b}}+p_{\Lambda }\right) _{\mu }\bar{%
u}_{\Lambda }u_{\Lambda _{b}}+i\bar{u}_{\Lambda }\sigma _{\mu \nu
}q^{\nu }u_{\Lambda _{b}},  \label{eq1} \\ \left( M_{\Lambda
}-M_{\Lambda _{b}}\right) \bar{u}_{\Lambda }\gamma _{\mu }\gamma
_{5}u_{\Lambda _{b}} &=&\left( p_{\Lambda _{b}}+p_{\Lambda
}\right) _{\mu }\bar{u}_{\Lambda }\gamma _{5}u_{\Lambda
_{b}}+i\bar{u}_{\Lambda }\sigma _{\mu \nu }q^{\nu }\gamma
_{5}u_{\Lambda _{b}}\,.  \label{eq2}
\end{eqnarray}
The form factors for dipole operators are derived as
\begin{eqnarray}
\left\langle \Lambda \right| \bar{s}i\sigma _{\mu \nu }q^{\nu }b\left|
\Lambda _{b}\right\rangle &=&f_{1}^{T}\bar{u}_{\Lambda }\gamma _{\mu
}u_{\Lambda _{b}}+f_{2}^{T}\bar{u}_{\Lambda }i\sigma _{\mu \nu }q^{\nu
}u_{\Lambda _{b}}+f_{3}^{T}q_{\mu }\bar{u}_{\Lambda }u_{\Lambda _{b}},
\label{tcq} \\
\left\langle \Lambda \right| \bar{s}i\sigma _{\mu \nu }q^{\nu }\gamma
_{5}b\left| \Lambda _{b}\right\rangle &=&g_{1}^{T}\bar{u}_{\Lambda }\gamma
_{\mu }\gamma _{5}u_{\Lambda _{b}}+g_{2}^{T}\bar{u}_{\Lambda }i\sigma _{\mu
\nu }q^{\nu }\gamma _{5}u_{\Lambda _{b}}+g_{3}^{T}q_{\mu }\bar{u}_{\Lambda
}\gamma _{5}u_{\Lambda _{b}}.  \label{atcq}
\end{eqnarray}
with
\begin{eqnarray}
f_{2}^{T} &=&f_{T}-f_{T}^{S}q^{2}\,,
\nn \\
f_{1}^{T} &=&\left[ f_{T}^{V}+f_{T}^{S}\left( M_{\Lambda }+M_{\Lambda
_{b}}\right) \right] q^{2}\,,
\nn \\
f_{1}^{T} &=&-\frac{q^{2}}{\left( M_{\Lambda _{b}}-M_{\Lambda }\right) }%
f_{3}^{T}\,,\nn \\
g_{2}^{T} &=&g_{T}-g_{T}^{S}q^{2}\,,\nn \\
g_{1}^{T} &=&\left[ g_{T}^{V}+g_{T}^{S}\left( M_{\Lambda }-M_{\Lambda
_{b}}\right) \right] q^{2}\,,\nn \\
g_{1}^{T} &=&\frac{q^{2}}{\left( M_{\Lambda _{b}}+M_{\Lambda }\right) }%
g_{3}^{T}\,.
\end{eqnarray}

 We now give the most general formulas by including the
right-handed coupling in the effective Hamiltonian with a complete set
of form factors. The free quark decay amplitudes for $b\rightarrow
sl^{+}l^{-}$ are given by

\begin{eqnarray}
{\cal H}\left( b\rightarrow sl^{+}l^{-}\right) &=&\frac{G_{F}\alpha _{em}}{%
\sqrt{2}\pi }V_{tb}V_{ts}^{*}\left[ \bar{s}\gamma ^{\mu }\left(
C_{9}^{L}P_{L}+C_{9}^{R}P_{R}\right) b\ \bar{l}\gamma _{\mu }l\right.
\nonumber \\
&&+\bar{s}\gamma ^{\mu }\left( C_{10}^{L}P_{L}+C_{10}^{R}P_{R}\right) b\
\bar{l}\gamma _{\mu }\gamma _{5}l  \nonumber \\
&&\left. -\frac{2m_{b}}{q^{2}}\bar{s}i\sigma _{\mu \nu }q^{\nu }\left(
C_{7}^{L}P_{R}+C_{7}^{R}P_{L}\right) b\ \bar{l}\gamma _{\mu }l\right]
\label{hameff}
\end{eqnarray}
where $C_{i}^{L}$ and $C_{i}^{R}\left( i=7,9,10\right) $ denote the
effective Wilson coefficients of left- and right-handed couplings,
respectively. With the most general form factors in Eqs.
(\ref{vc}), (\ref{ac}), (\ref{tcq}) and (\ref{atcq}), and the effective
Hamiltonian in Eq.
(\ref{hameff}), the transition matrix elements for
the decays of $\Lambda _{b}\rightarrow
\Lambda l^{+}l^{-}$  are expressed as
\begin{eqnarray}
{\cal M}\left( \Lambda _{b}\rightarrow \Lambda l^{+}l^{-}\right) &=&\frac{%
G_{F}\alpha _{em}}{\sqrt{2}}V_{tb}V_{ts}^{*}\left\{ \left[ \bar{\Lambda}%
\gamma _{\mu }\left( A_{1}P_{R}+B_{1}P_{L}\right) \Lambda _{b}\right. \right.
\nonumber \\
&&\left. +\bar{\Lambda}i\sigma _{\mu \nu }q^{\nu }\left(
A_{2}P_{R}+B_{2}P_{L}\right) \Lambda _{b}\right] \ \bar{l}\gamma _{\mu }l
\nonumber \\
&&+\left[ \bar{\Lambda}\gamma _{\mu }\left( D_{1}P_{R}+E_{1}P_{L}\right)
\Lambda _{b}+\bar{\Lambda}i\sigma _{\mu \nu }q^{\nu }\left(
D_{2}P_{R}+E_{2}P_{L}\right) \Lambda _{b}\right.  \nonumber \\
&&\left. \left. +q_{\mu }\bar{\Lambda}\left( D_{3}P_{R}+E_{3}P_{L}\right)
\Lambda _{b}\right] \ \bar{l}\gamma _{\mu }\gamma _{5}l\right\}
 \label{eff}
\end{eqnarray}
where

\begin{eqnarray}
A_{i} &=&C_{9}^{R}\frac{f_{i}+g_{i}}{2}-\frac{2m_{b}}{q^{2}}C_{7}^{R}\frac{%
f_{i}^{T}-g_{i}^{T}}{2}+C_{9}^{L}\frac{f_{i}-g_{i}}{2}-\frac{2m_{b}}{q^{2}}%
C_{7}^{L}\frac{f_{i}^{T}{}+g_{i}^{T}}{2},  \nonumber \\
B_{i} &=&C_{9}^{L}\frac{f_{i}+g_{i}}{2}-\frac{2m_{b}}{q^{2}}C_{7}^{L}\frac{%
f_{i}^{T}-g_{i}^{T}}{2}+C_{9}^{R}\frac{f_{i}-g_{i}}{2}-\frac{2m_{b}}{q^{2}}%
C_{7}^{R}\frac{f_{i}^{T}{}+g_{i}^{T}}{2},  \nonumber \\
D_{i} &=&C_{10}^{R}\frac{f_{i}+g_{i}}{2}+C_{10}^{L}\frac{f_{i}-g_{i}}{2},
\nonumber \\
E_{i} &=&C_{10}^{L}\frac{f_{i}+g_{i}}{2}+C_{10}^{R}\frac{f_{i}-g_{i}}{2}.
\\
\nn
\label{ceff}
\end{eqnarray}

\noindent
{\large
\bf B. Differential decay rates}\\

Using the transition matrix elements in Eq. (\ref{eff}), the double
differential decay rates can be derived as
\be
\frac{d\Gamma }{dsd\hat{z}} &=&\frac{G_{F}^{2}\alpha _{em}^{2}\lambda
_{t}^{2}}{768\pi ^{5}}M_{\Lambda _{b}}^{5}\sqrt{\phi \left( s\right) }\sqrt{%
1-\frac{4m_{l}^{2}}{q^{2}}}\bar{f}^{2}R_{\Lambda _{b}}\left( s,\hat{z}%
\right) \,,
\label{apprate}
\ee
where
\be
R_{\Lambda _{b}}\left( s,\hat{z}\right) &=&I_{0}\left( s,\hat{z}\right) +%
\hat{z}I_{1}\left( s,\hat{z}\right) +\hat{z}^{2}I_{2}\left( s,\hat{z}\right)
\ee
with

\begin{eqnarray}
I_{0}\left( s,\hat{z}\right) &=&-6\sqrt{r}\hat{s}\left[ \left( 1+2\frac{%
m_{l}^{2}}{q^{2}}\right) {\rm {Re}A_{1}B_{1}^{*}+\left( 1-6\frac{m_{l}^{2}}{%
q^{2}}\right) {Re}D_{1}E_{1}^{*}}\right]
\nn\\
&&+\frac{3}{4}\left( \left( 1-r\right) ^{2}-s^{2}\right) \left( \left|
A_{1}\right| ^{2}+\left| B_{1}\right| ^{2}+\left| D_{1}\right| ^{2}+\left|
E_{1}\right| ^{2}\right)
\nn\\
&&+6\hat{m}_{l}^{2}t\left( \left| A_{1}\right| ^{2}+\left| B_{1}\right|
^{2}-\left| D_{1}\right| ^{2}-\left| E_{1}\right| ^{2}\right)
\nn\\
&&+12\hat{m}_{l}^{2}M_{\Lambda _{b}}\sqrt{r}\left( 1-t\right) \left( {\rm {Re%
}D_{1}D_{3}^{\prime *}+{Re}E_{1}E_{3}^{\prime *}}\right)
\nn\\
&&+12\hat{m}_{l}^{2}M_{\Lambda _{b}}\left( t-r\right) \left( {\rm {Re}%
D_{1}E_{3}^{\prime *}+{Re}D_{3}E_{1}^{\prime *}}\right)
\nn\\
&&+6M_{\Lambda _{b}}\sqrt{r}s\left( 1-t\right) \left[ \left( 1+2\frac{%
m_{l}^{2}}{q^{2}}\right) \left( {\rm {Re}A_{1}A_{2}^{*}+{Re}B_{1}B_{2}^{*}}%
\right) \right.
\nn \\
&&\left. +\left( 1-2\frac{m_{l}^{2}}{q^{2}}\right) \left( {\rm {Re}%
D_{1}D_{2}^{*}+{Re}E_{1}E_{2}^{*}}\right) \right] \nn\\
&&-6M_{\Lambda _{b}}s\left( t-r\right) \left[ \left( 1+2\frac{m_{l}^{2}}{%
q^{2}}\right) \left( {\rm {Re}A_{1}B_{2}^{*}+{Re}A_{2}B_{1}^{*}}\right)
\right. \nn\\
&&\left. +\left( 1-6\frac{m_{l}^{2}}{q^{2}}\right) \left( {\rm {Re}%
D_{1}E_{2}^{*}+{Re}D_{2}E_{1}^{*}}\right) \right] \nn\\
&&-6M_{\Lambda _{b}}^{2}\sqrt{r}s^{2}\left( 1+2\frac{m_{l}^{2}}{q^{2}}%
\right) {\rm {Re}A_{2}B_{2}^{*}-6M_{\Lambda _{b}}^{2}\sqrt{r}s^{2}\left( 1-6%
\frac{m_{l}^{2}}{q^{2}}\right) {Re}D_{2}E_{2}^{*}} \nn\\
&&-6M_{\Lambda _{b}}^{2}\left[ s\left( 1-t\right) \left( t-r\right) -\frac{1%
}{8}\left( \left( 1-r\right) ^{2}-s^{2}\right) \right] \left( \left|
A_{2}\right| ^{2}+\left| B_{2}\right| ^{2}+\left| D_{2}\right| ^{2}+\left|
E_{2}\right| ^{2}\right) \nn\\
&&-6M_{\Lambda _{b}}^{2}\hat{m}_{l}^{2}\left( 2r-\left( 1+r\right) t\right)
\left( \left| A_{2}\right| ^{2}+\left| B_{2}\right| ^{2}-\left| D_{2}\right|
^{2}-\left| E_{2}\right| ^{2}\right) \nn\\
&&+12\hat{m}_{l}^{2}M_{\Lambda _{b}}^{2}st\left( {\rm {Re}D_{2}D_{3}^{\prime
*}+{Re}E_{2}E_{3}^{\prime *}}\right) +12\hat{m}_{l}^{2}M_{\Lambda _{b}}^{2}%
\sqrt{r}s\left( {\rm {Re}D_{2}E_{3}^{\prime *}+{Re}D_{3}^{\prime *}E_{2}}%
\right) \,,
\nn\\
I_{1}\left( s,\hat{z}\right) &=&3s\phi \left( s\right) \left\{ -\left( {\rm {%
Re}A_{1}D_{1}^{*}-{Re}B_{1}E_{1}^{*}}\right) +M_{\Lambda _{b}}\left[ \sqrt{r}%
\left( {\rm {Re}A_{1}D_{2}^{*}-{Re}B_{1}E_{2}^{*}}\right) \right. \right.
\nn\\
&&\left. +\left( {\rm {Re}A_{1}E_{2}^{*}-{Re}B_{1}D_{2}^{*}}\right) +\sqrt{r}%
\left( {\rm {Re}A_{2}D_{1}^{*}-{Re}B_{2}E_{1}^{*}}\right) -\left( {\rm {Re}%
A_{2}E_{1}^{*}-{Re}B_{2}D_{1}^{*}}\right) \right]
\nn\\
&&\left. +M_{\Lambda _{b}}\left( 1-r\right) \left( {\rm {Re}A_{2}D_{2}^{*}-{%
Re}B_{2}E_{2}^{*}}\right) \right\} \,,
\nn\\
I_{2}\left( s,\hat{z}\right) &=&-\frac{3}{4}\phi \left( s\right) \left( 1-4%
\frac{m_{l}^{2}}{q^{2}}\right) \left( \left| A_{1}\right| ^{2}+\left|
B_{1}\right| ^{2}+\left| D_{1}\right| ^{2}+\left| E_{1}\right| ^{2}\right)
\nn\\
&&+\frac{3}{4}M_{\Lambda _{b}}^{2}\phi \left( s\right) \left( 1-4\frac{%
m_{l}^{2}}{q^{2}}\right) \left( \left| A_{2}\right| ^{2}+\left| B_{2}\right|
^{2}+\left| D_{2}\right| ^{2}+\left| E_{2}\right| ^{2}\right)
\end{eqnarray}
where $D_{3}^{\prime }=D_{3}-D_{2}$ and $E_{3}^{\prime }=E_{3}-E_{2}$, and
$\hat{z}=\hat{p}_{B}\cdot \hat{p}_{l^{+}}$ denotes the angle between the
momentum of $\Lambda _{b}$ and that of $l^{+}$ in the di-lepton
invariant mass frame.\\

\noindent
{\large \bf C. Forward-backward and lepton asymmetries}
\\

 From Eq. (\ref{eff}), the functions of
$R_{\Lambda_b}$ and $R_{FB}$ in the
 differential and normalized FBAs
for $\Lambda _{b}\to \Lambda l^{+}l^{-}$ in Eqs. (\ref{dfba}) and
(\ref{fba}) are given by

\begin{eqnarray}
R_{\Lambda _{b}}\left( s\right) &=&\frac{1}{2}\int_{-1}^{1}d\hat{z}%
R_{\Lambda _{b}}\left( s,\hat{z}\right)
\ee
and
\be
R_{FB}\left( s\right) &=&-{\rm {Re}\left(
A_{1}D_{1}^{*}-B_{1}E_{1}^{*}\right) +M_{\Lambda _{b}}\left[ \sqrt{r}{Re}%
\left( A_{1}D_{2}^{*}-B_{1}E_{2}^{*}\right) +{Re}\left(
A_{1}E_{2}^{*}-B_{1}D_{2}^{*}\right) \right. }
\nn\\
&&\left. +\sqrt{r}{\rm {Re}\left( A_{2}D_{1}^{*}-B_{2}E_{1}^{*}\right) -{Re}%
\left( A_{2}E_{1}^{*}-B_{2}D_{1}^{*}\right) }\right]
\nn\\
&&+M_{\Lambda _{b}}^{2}(1-r){\rm {Re}\left(
A_{2}D_{2}^{*}-B_{2}E_{2}^{*}\right) }\,,
\end{eqnarray}
respectively.
We can also define the longitudinal and transverse
lepton polarization asymmetries.
 Explicitly,
 by the definition of Eq. (\ref{pasy}), we get
\begin{eqnarray}
P_{L} &=&-\frac{1}{R_{\Lambda _{b}}(s)}\sqrt{1-\frac{4\hat{m}_{l}^{2}}{s}}%
\left\{ \left( s\left( 1+r-s\right) +(1-r)^{2}-s^{2}\right) {\rm {Re}\left(
A_{1}D_{1}^{*}+B_{1}E_{1}^{*}\right) }\right.
\nn\\
&&+M_{\Lambda _{b}}^{2}s\left( -s\left( 1+r-s\right)
+2(1-r)^{2}-2s^{2}\right) {\rm {Re}\left(
A_{2}D_{2}^{*}+B_{2}E_{2}^{*}\right) }
\nn\\
&&-6s\sqrt{r}\left[ {\rm {Re}\left( A_{1}E_{1}^{*}+B_{1}D_{1}^{*}\right)
+M_{\Lambda _{b}}^{2}s{Re}\left( A_{2}E_{2}^{*}+B_{2}D_{2}^{*}\right) }%
\right]
\nn\\
&&+3sM_{\Lambda }\left( 1-r+s\right) \left[ {\rm {Re}\left(
A_{1}D_{2}^{*}+B_{1}E_{2}^{*}\right) +{Re}\left(
A_{2}D_{1}^{*}+B_{2}E_{1}^{*}\right) }\right] \nn\\
&&-3sM_{\Lambda _{b}}\left( 1-r-s\right) \left[ {\rm {Re}\left(
A_{1}E_{2}^{*}+B_{1}D_{2}^{*}\right) -}+{\rm {e}\left(
A_{2}E_{1}^{*}+B_{2}D_{1}^{*}\right) }\right] , \\
P_{T} &=&\frac{3}{4}\pi \hat{m}_{l}\sqrt{1-\frac{4m_{l}^{2}}{q^{2}}}\sqrt{%
s\phi \left( s\right) }\frac{1}{R_{\Lambda _{b}}(s)}\left\{ -{\rm {Im}\left(
A_{1}D_{1}^{*}-B_{1}E_{1}^{*}\right) }\right. \nn\\
&&+M_{\Lambda }\left[ {\rm {Im}\left( A_{1}D_{2}^{*}-B_{1}E_{2}^{*}\right) +{%
Im}\left( A_{2}D_{1}^{*}-B_{2}E_{1}^{*}\right) }\right] \nn\\
&&+M_{\Lambda _{b}}\left[ {\rm {Im}\left(
A_{1}E_{2}^{*}-B_{1}D_{2}^{*}\right) -{Im}\left(
A_{2}E_{1}^{*}-B_{2}D_{1}^{*}\right) }\right] \nn\\
&&+M_{\Lambda _{b}}^{2}\left( 1-r\right) {\rm {Im}\left(
A_{2}D_{2}^{*}-B_{2}E_{2}^{*}\right) .}
\end{eqnarray}

\newpage
\begin{figcap}
\item
BRs as a function of $q^2/M^2_{\Lambda_b}$ for (a) $%
\Lambda_b\to\Lambda\mu^+\mu^-$ and (b) $\Lambda_b\to\Lambda\tau^+\tau^-$.
The curves with and without resonant shapes represent including and no LD
contributions, respectively. The dashed, solid and dash-dotted curves stand
for $\omega=0.15$, $0$, and $-0.15$, respectively.

\item
Same as Figure 1 but for the FBAs.

\item
Same as Figure 1 but for the
 longitudinal polarization asymmetries.

\item
Same as Figure 1 but for the
transverse polarization asymmetries.

\item
FBAs in the generic SUSY model as a function of
$q^2/M^2_{\Lambda_b}$
for (a) $\Lambda_b\to\Lambda\mu^+\mu^-$ and (b) $\Lambda_b\to\Lambda\tau^+%
\tau^-$. The solid and dashed curves stand for the SM and SUSY model,
respectively.

\item
Same as Figure 5 but for the
 longitudinal polarization asymmetries.

\item
Same as Figure 5 but for the
transverse polarization asymmetries.

\item
Same as Figure 5 but for
$\Delta_\Gamma$.

\item
Same as Figure 5 but for
$\Delta_{FBA}$.

\item
Same as Figure 5 but for
$\Delta_{P_L}$.

\item
Same as Figure 5 but for
$\Delta_{P_T}$.

\end{figcap}

\newpage
\begin{figure}[h]
\includegraphics{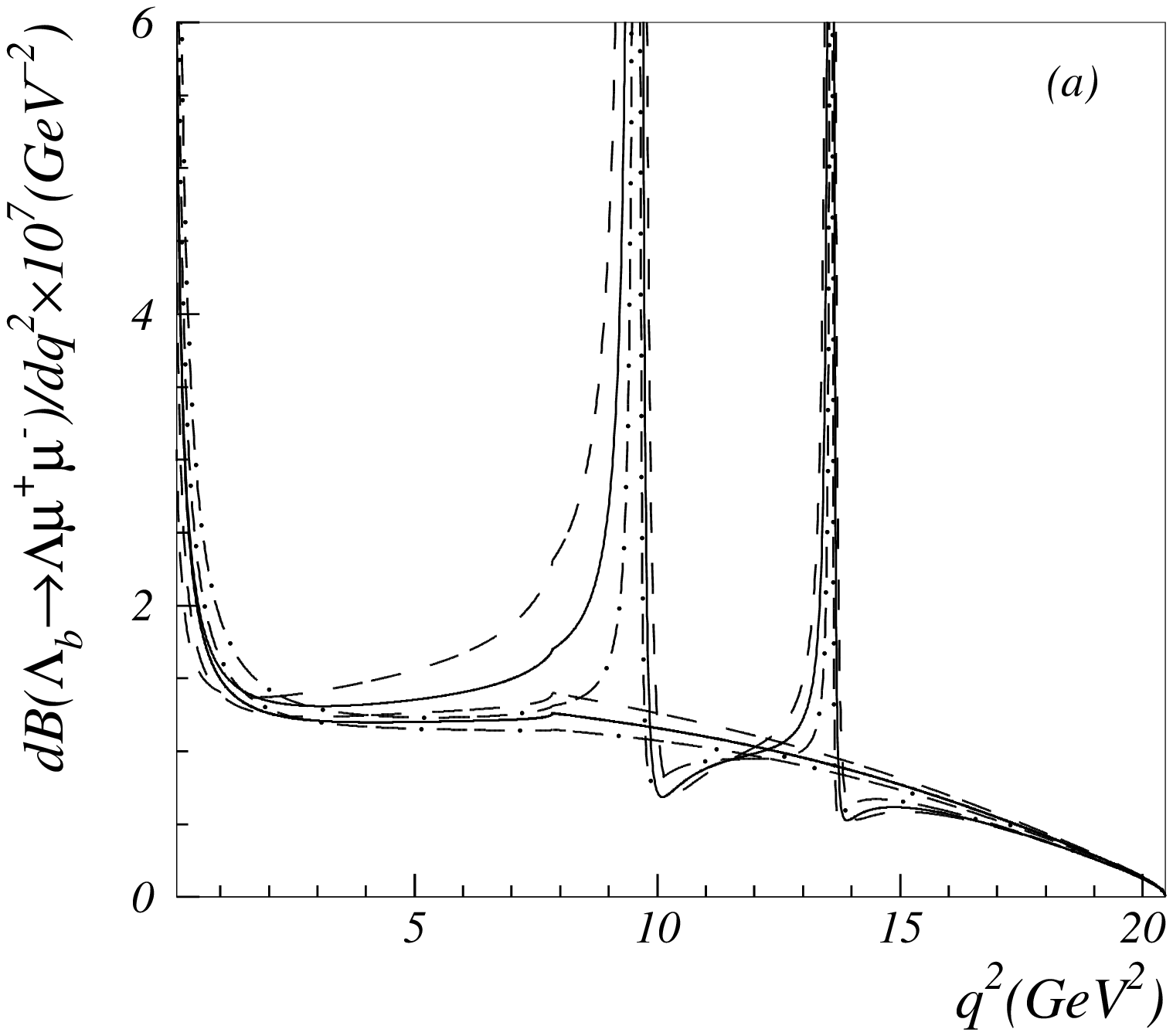}
\vskip 5.5cm
\end{figure}

\vskip 2.cm
\begin{figure}[h]
\includegraphics{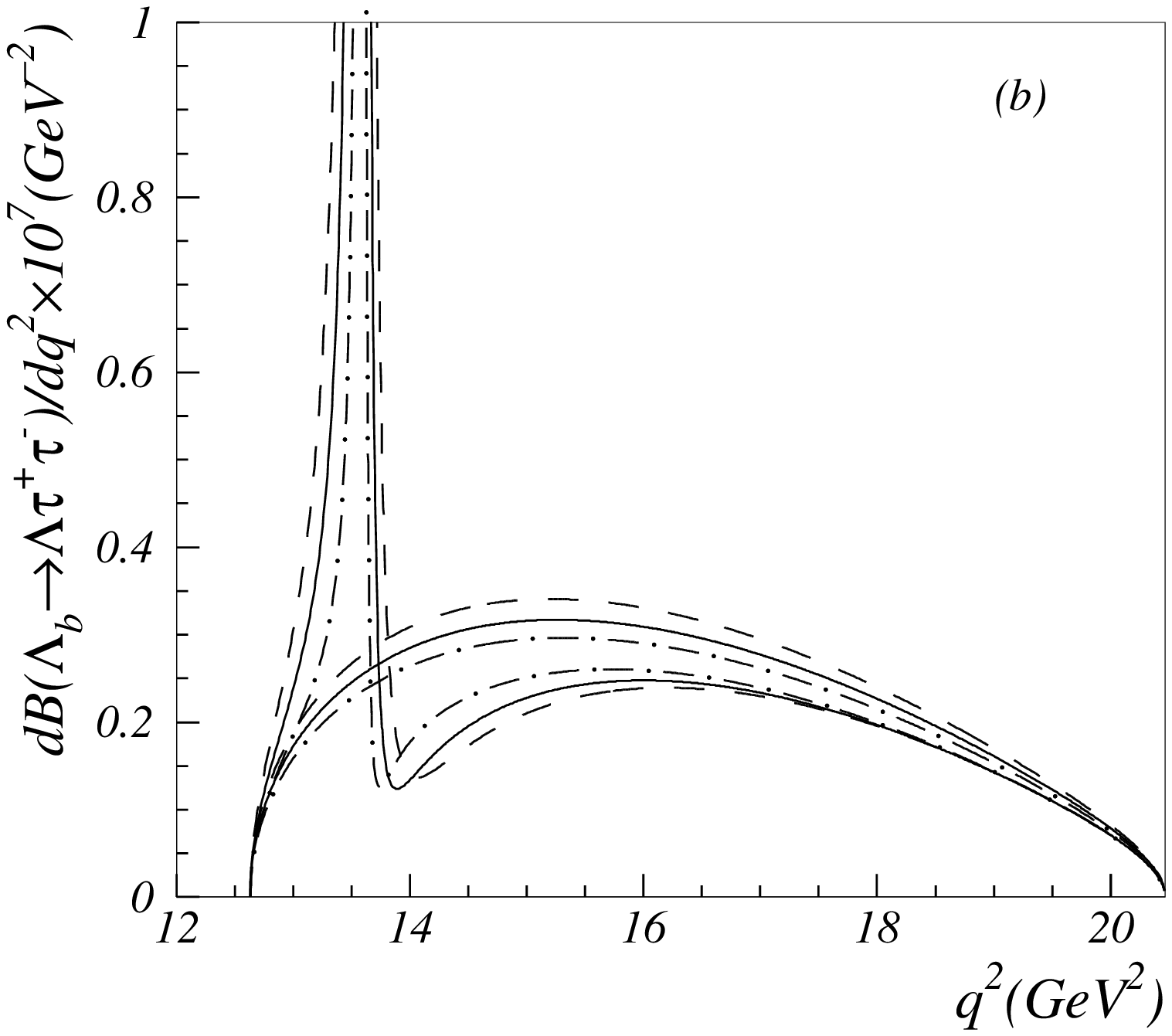}
\vskip 8.cm
\caption{ BRs as a function of $q^2/M^2_{\Lambda_b}$ for (a) $%
\Lambda_b\to\Lambda\mu^+\mu^-$ and (b) $\Lambda_b\to\Lambda\tau^+\tau^-$.
The curves with and without resonant shapes represent including and no LD
contributions, respectively. The dashed, solid and dash-dotted curves stand
for $\omega=0.15$, $0$, and $-0.15$, respectively. }
\end{figure}

\newpage
\begin{figure}[h]
\includegraphics{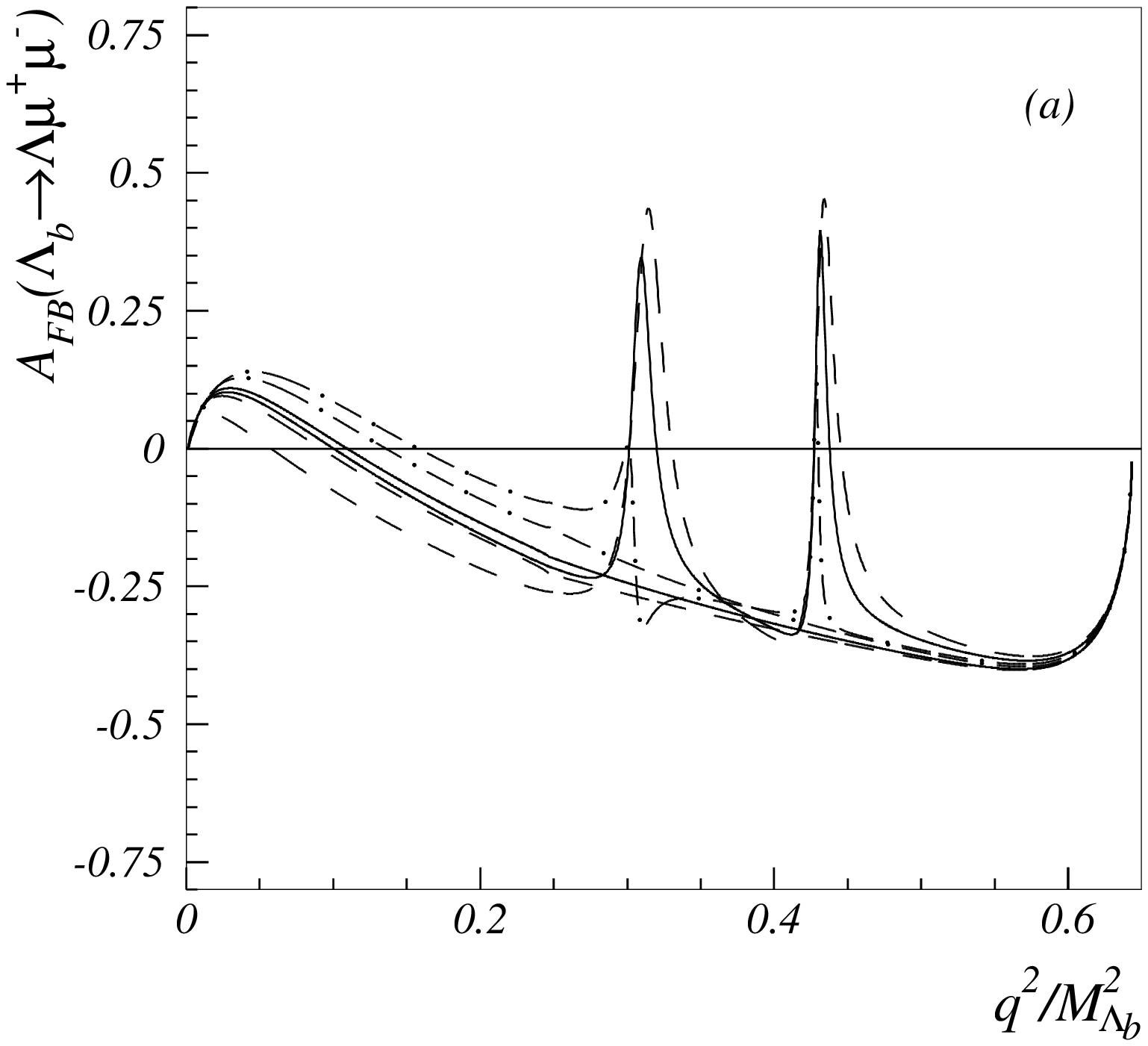}
\vskip 5.5cm
\end{figure}

\vskip 2.cm
\begin{figure}[h]
\includegraphics{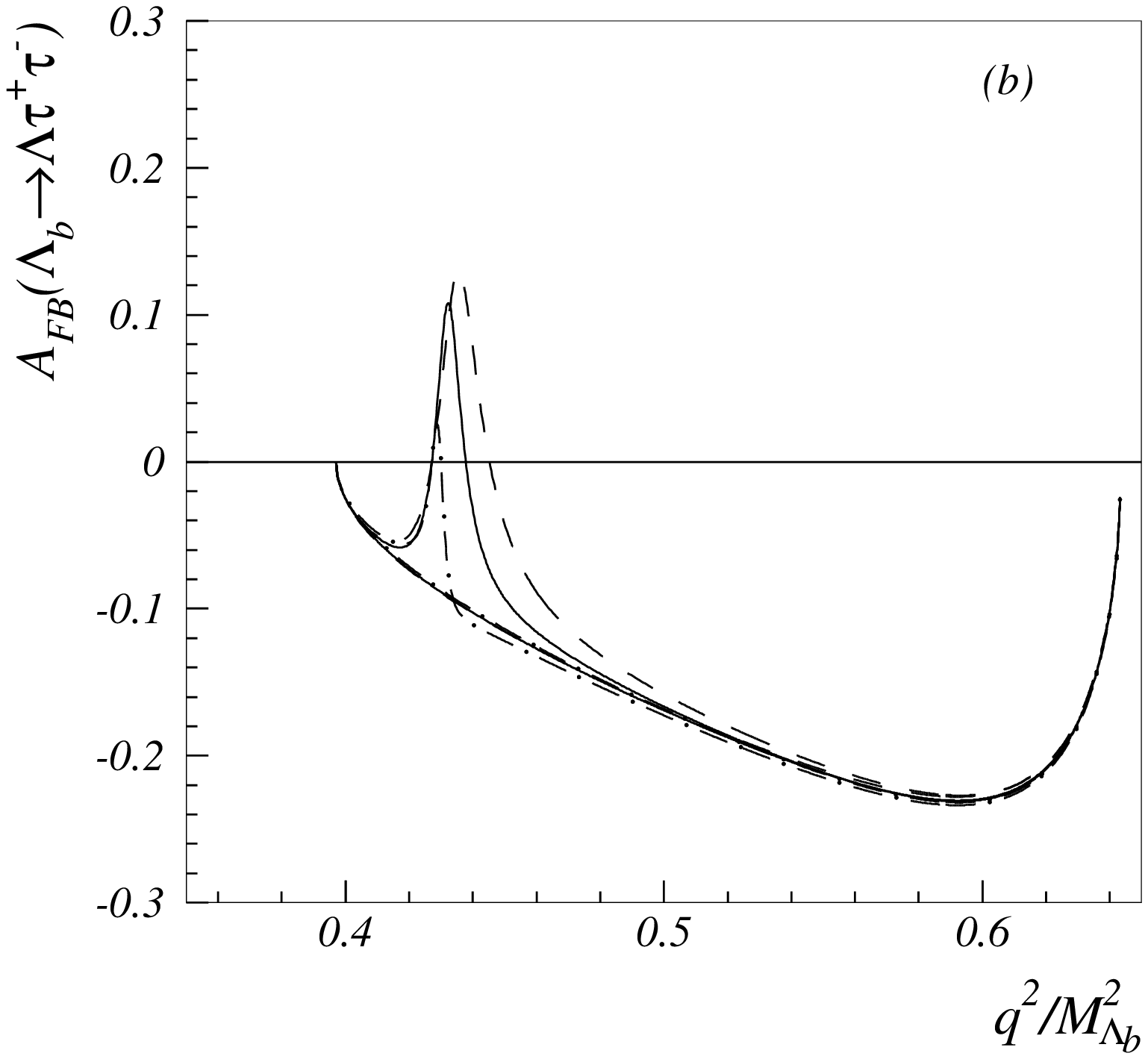}
\vskip 8.cm
\caption{
Same as Figure 1 but for the FBAs.}
\end{figure}

\newpage
\begin{figure}[h]
\includegraphics{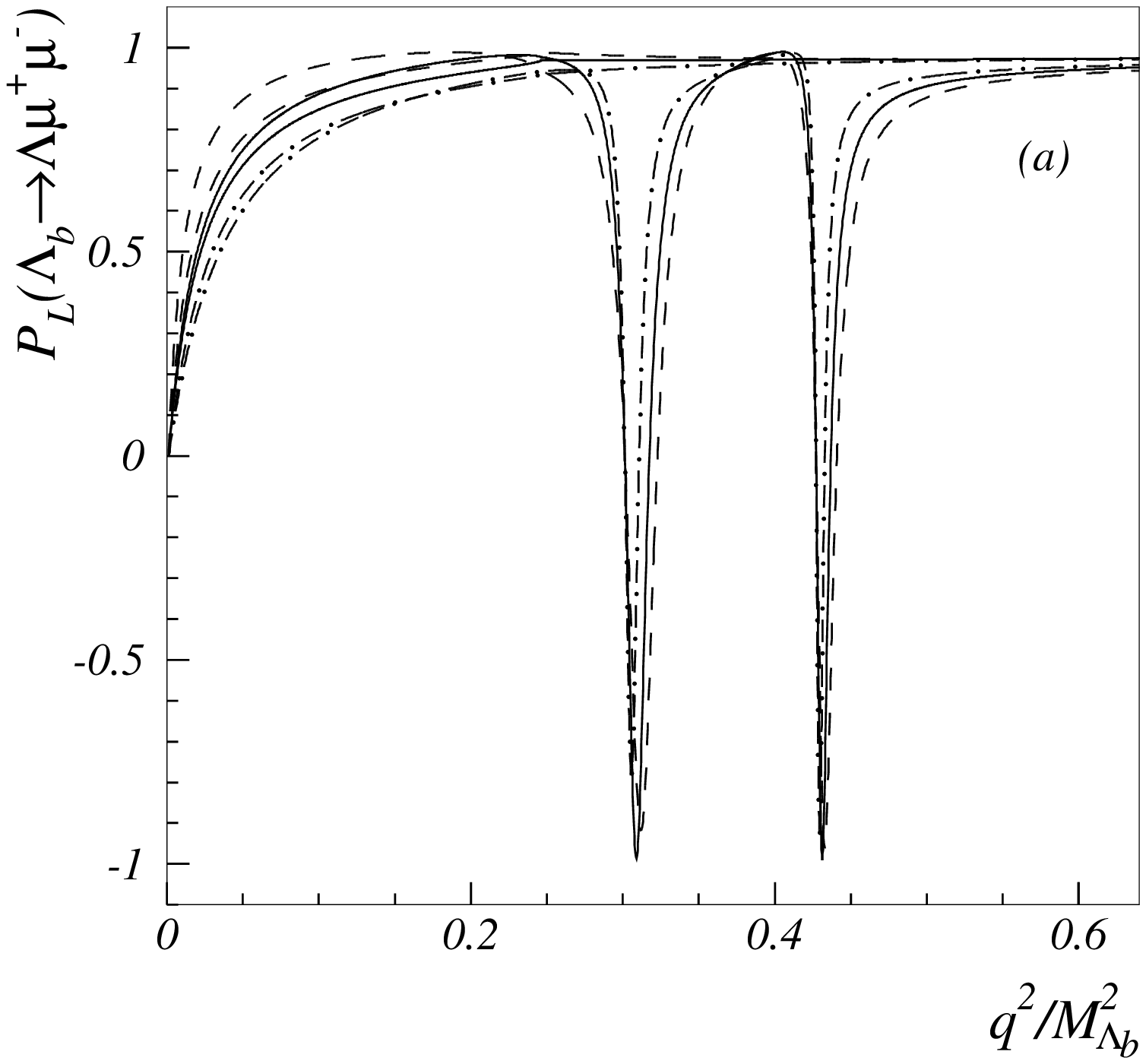}
\vskip 5.5cm
\end{figure}

\vskip 2.cm
\begin{figure}[h]
\includegraphics{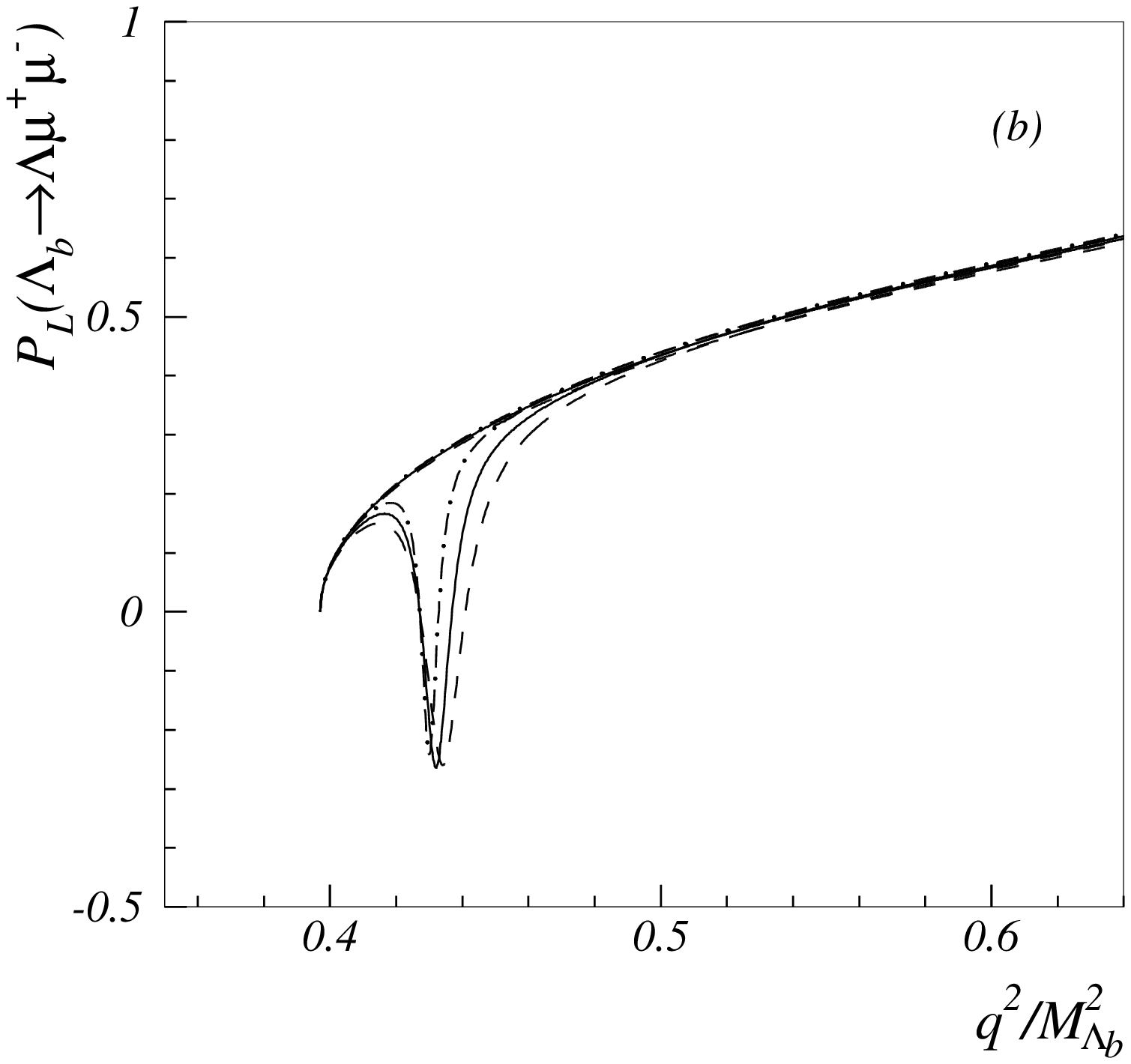}
\vskip 8.cm
\caption{
Same as Figure 1 but for the
 longitudinal polarization asymmetries.}
\end{figure}

\newpage
\begin{figure}[h]
\includegraphics{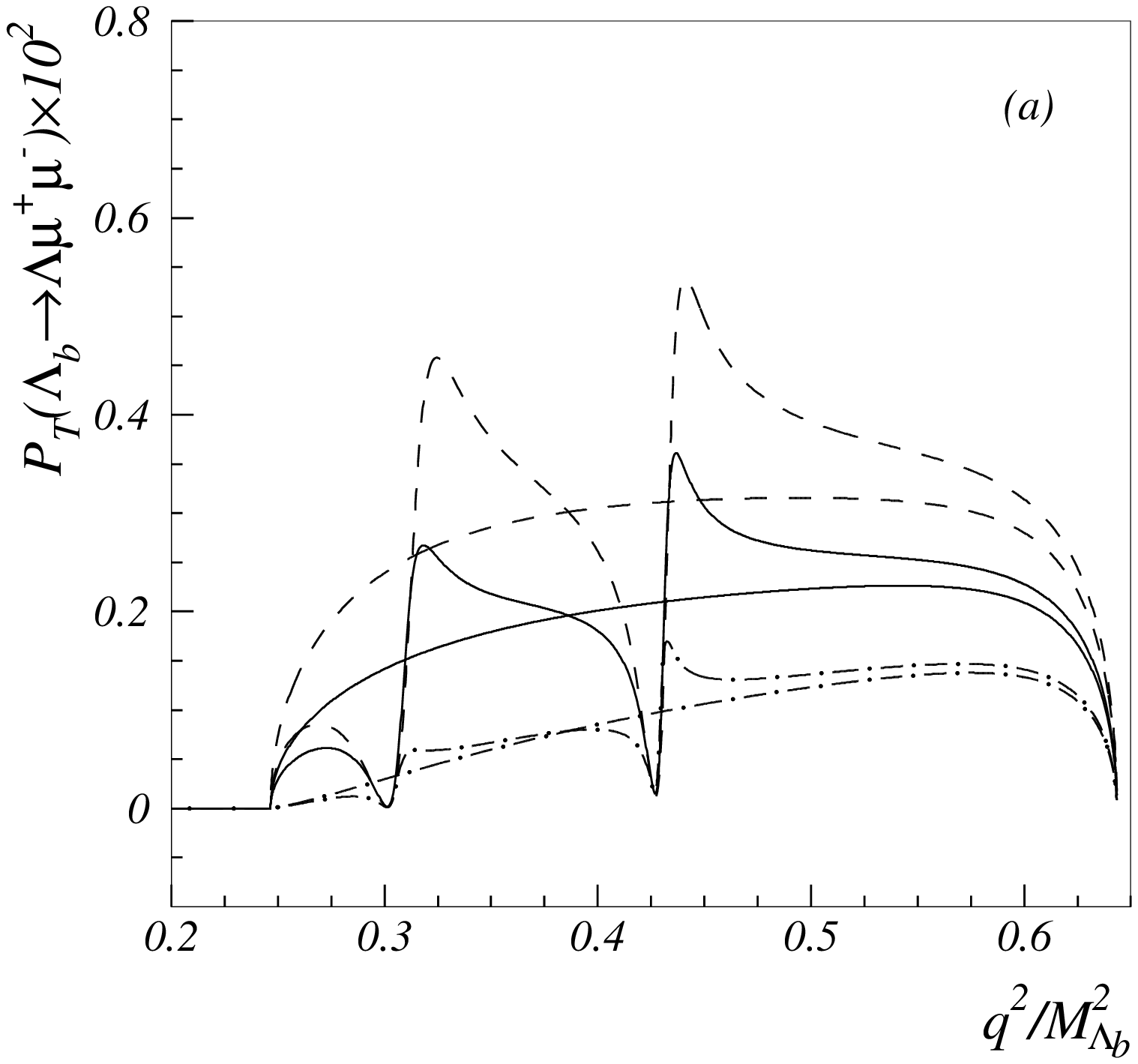}
\vskip 5.5cm
\end{figure}

\vskip 2.cm
\begin{figure}[h]
\includegraphics{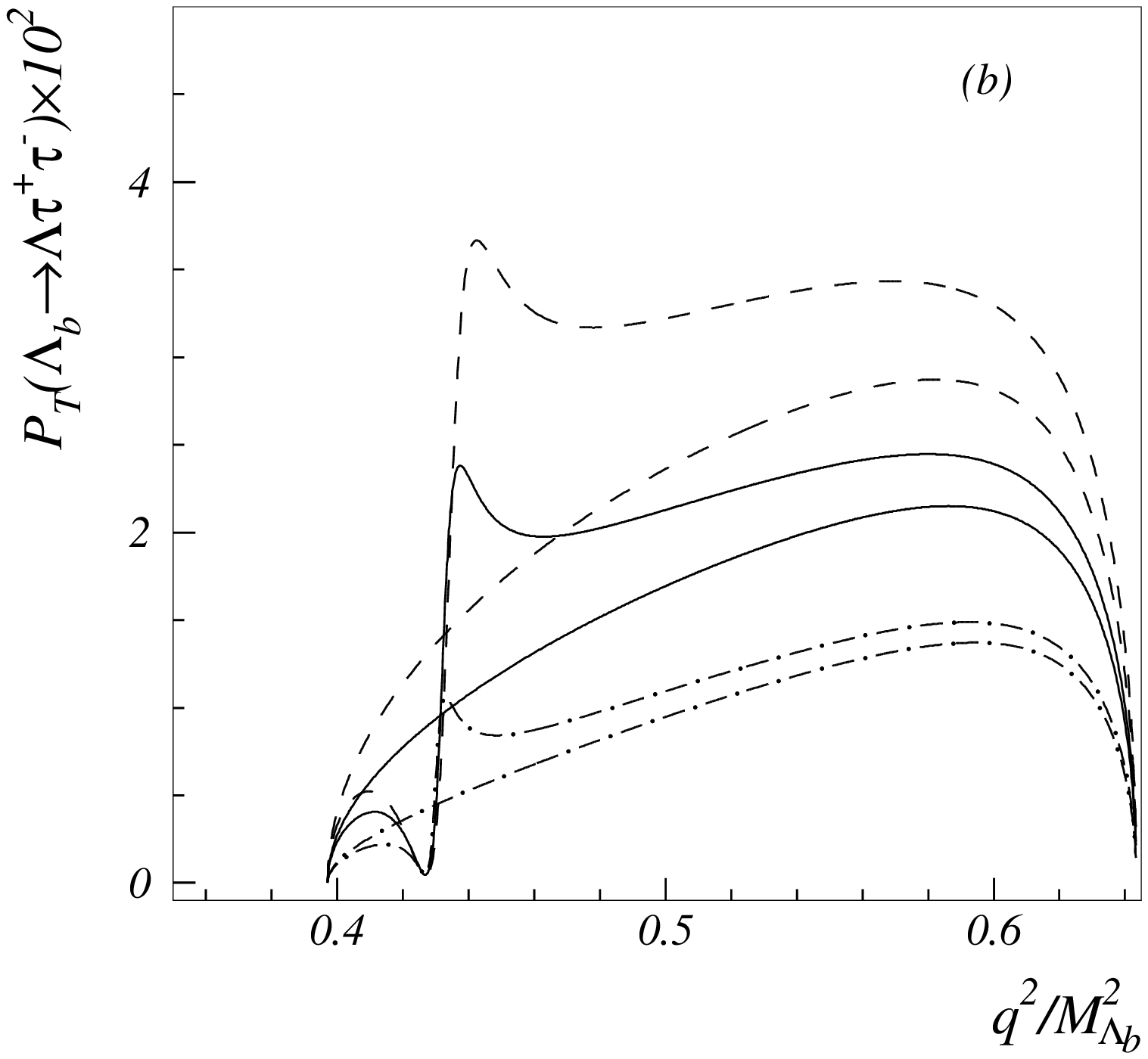}
\vskip 8.cm
\caption{
Same as Figure 1 but for the
transverse polarization asymmetries.}
\end{figure}

\newpage
\begin{figure}[h]
\includegraphics{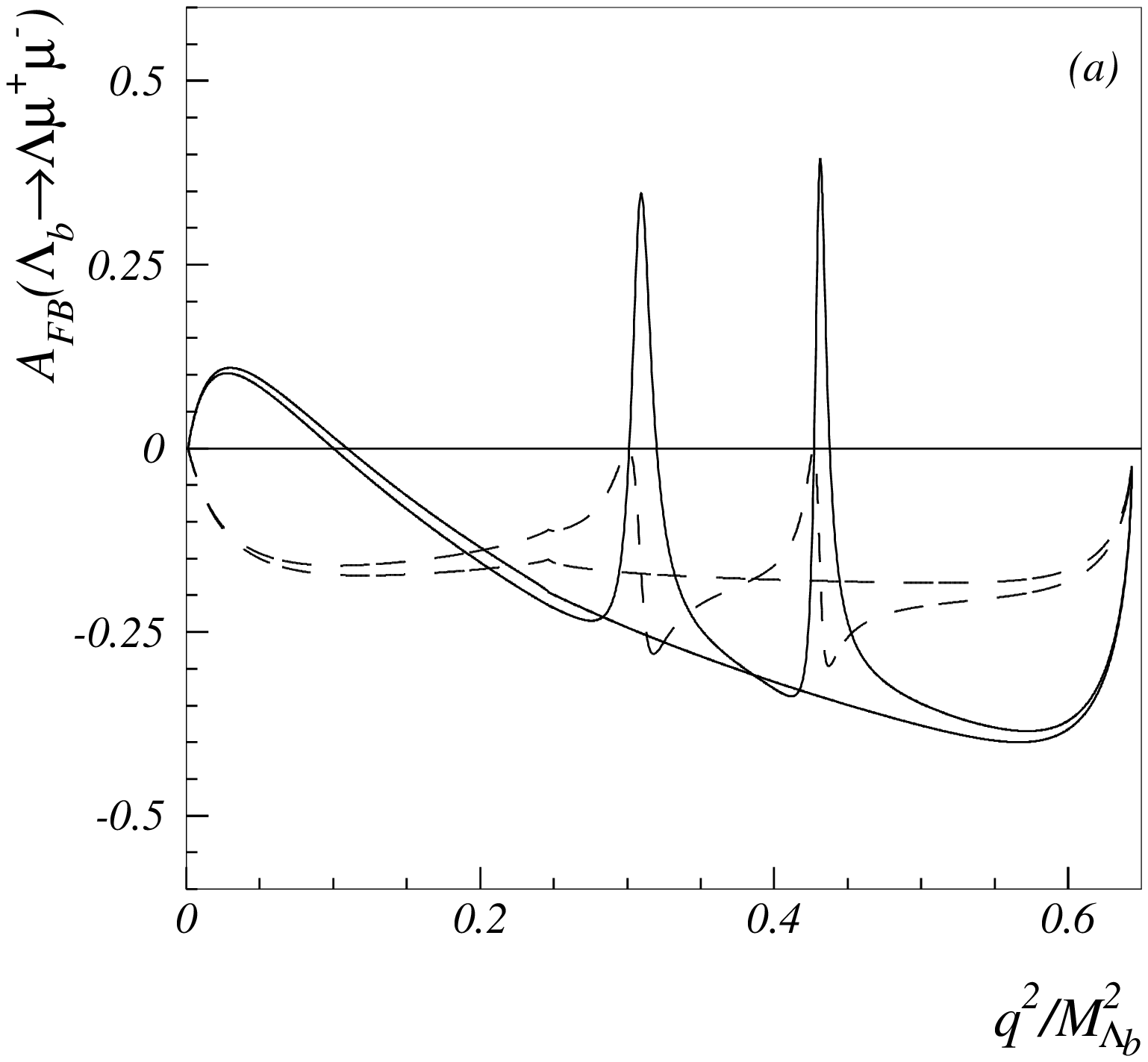}
\vskip 5.5cm
\end{figure}

\vskip 2.cm
\begin{figure}[h]
\includegraphics{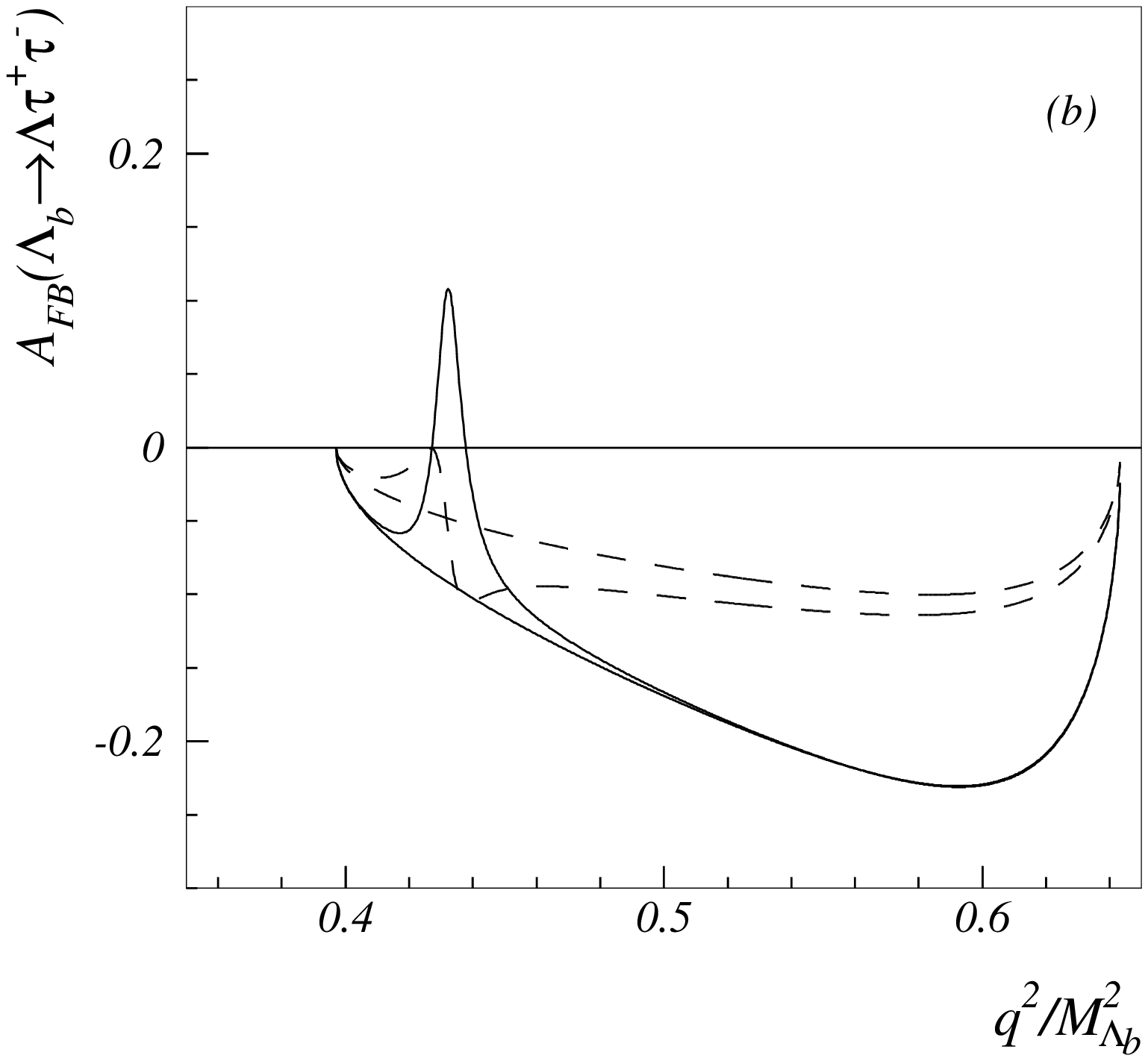}
\vskip 8.cm
\caption{ FBAs in the generic SUSY model as a function of
$q^2/M^2_{\Lambda_b}$
for (a) $\Lambda_b\to\Lambda\mu^+\mu^-$ and (b) $\Lambda_b\to\Lambda\tau^+%
\tau^-$. The solid and dashed curves stand for the SM and SUSY model,
respectively. }
\end{figure}

\newpage
\begin{figure}[h]
\includegraphics{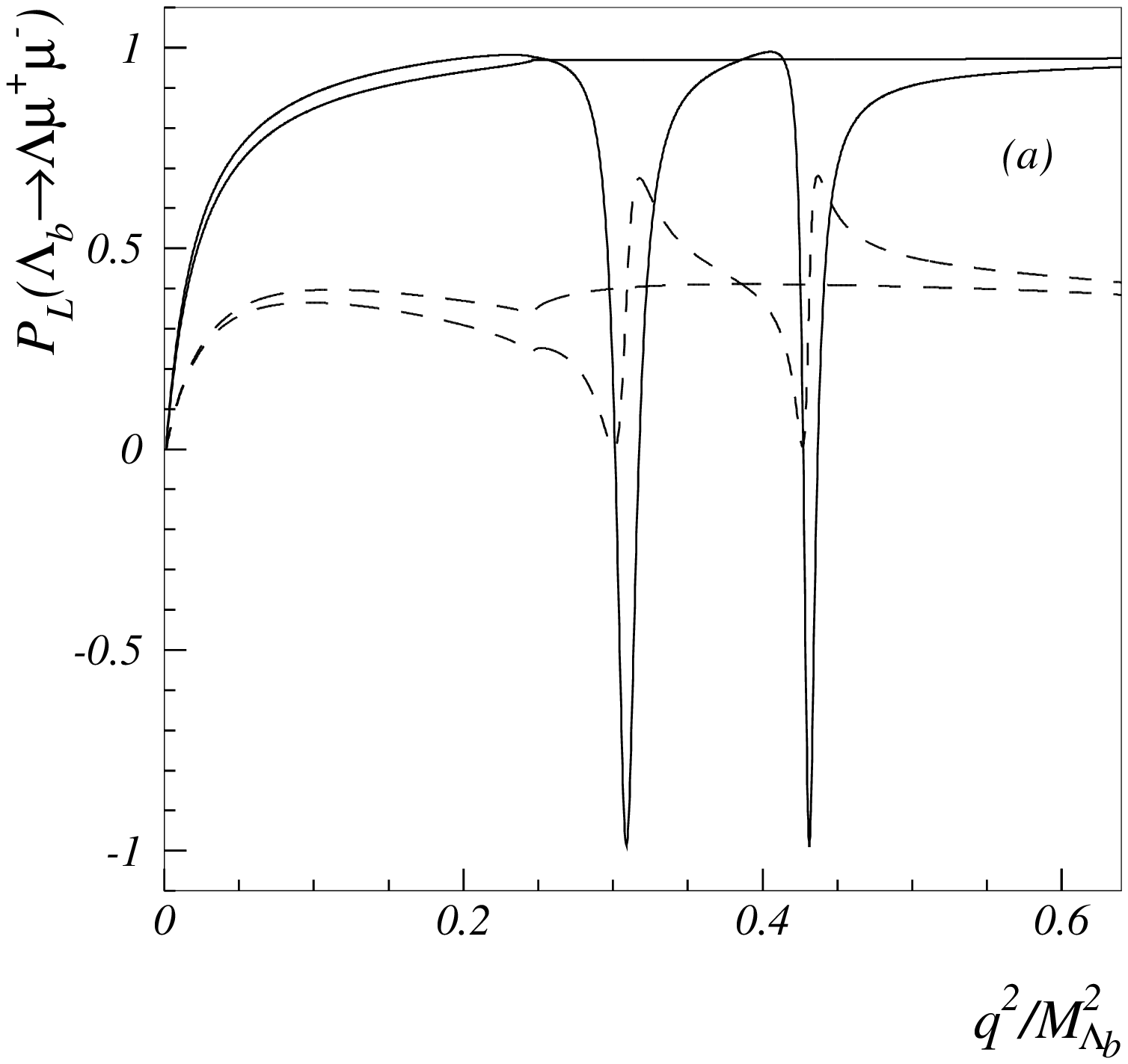}
\vskip 5.5cm
\end{figure}

\vskip 2.cm
\begin{figure}[h]
\includegraphics{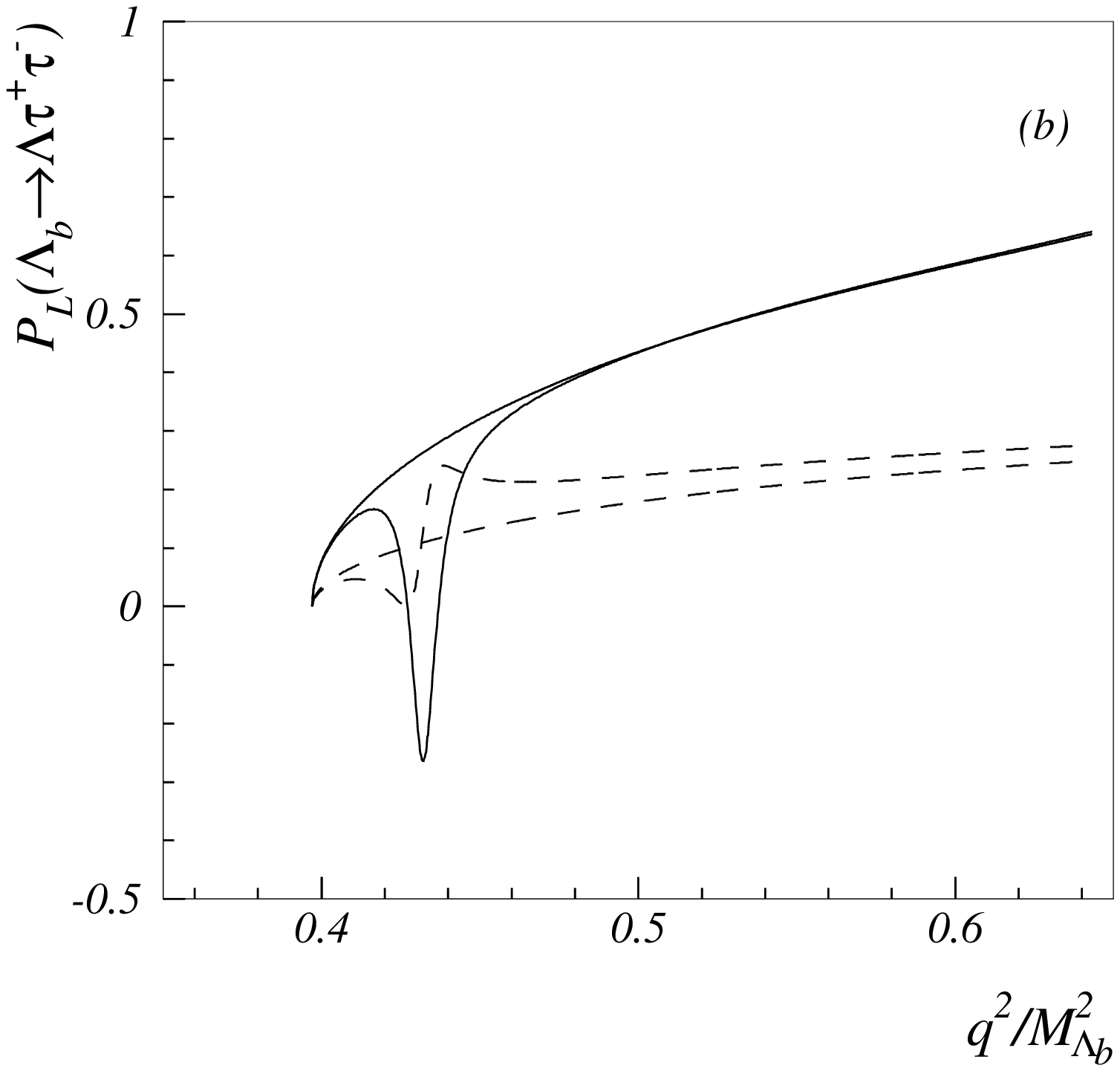}
\vskip 8.cm
\caption{
Same as Figure 5 but for the
 longitudinal polarization asymmetries.}
\end{figure}

\newpage
\begin{figure}[h]
\includegraphics{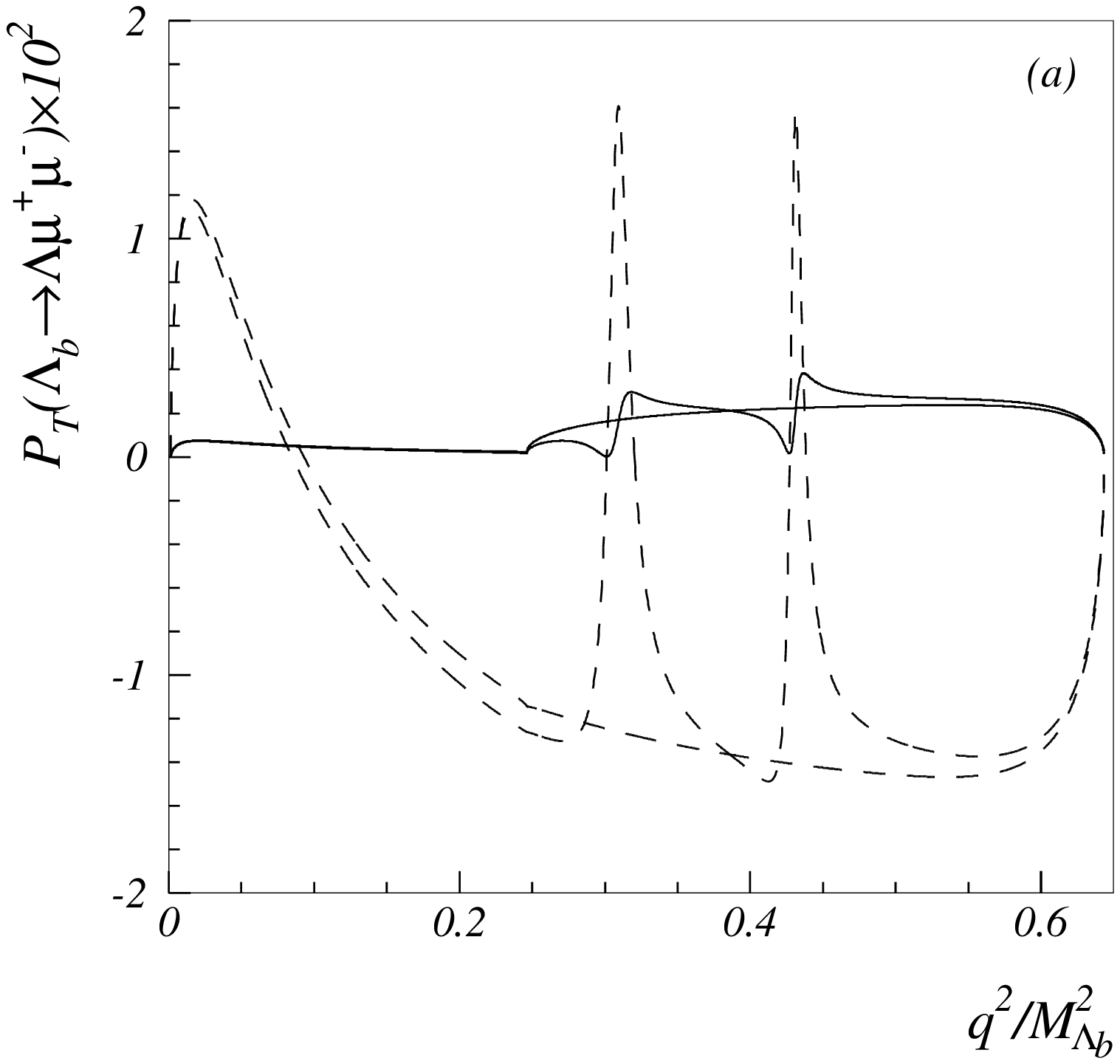}
\vskip 5.5cm
\end{figure}

\vskip 2.cm
\begin{figure}[h]
\includegraphics{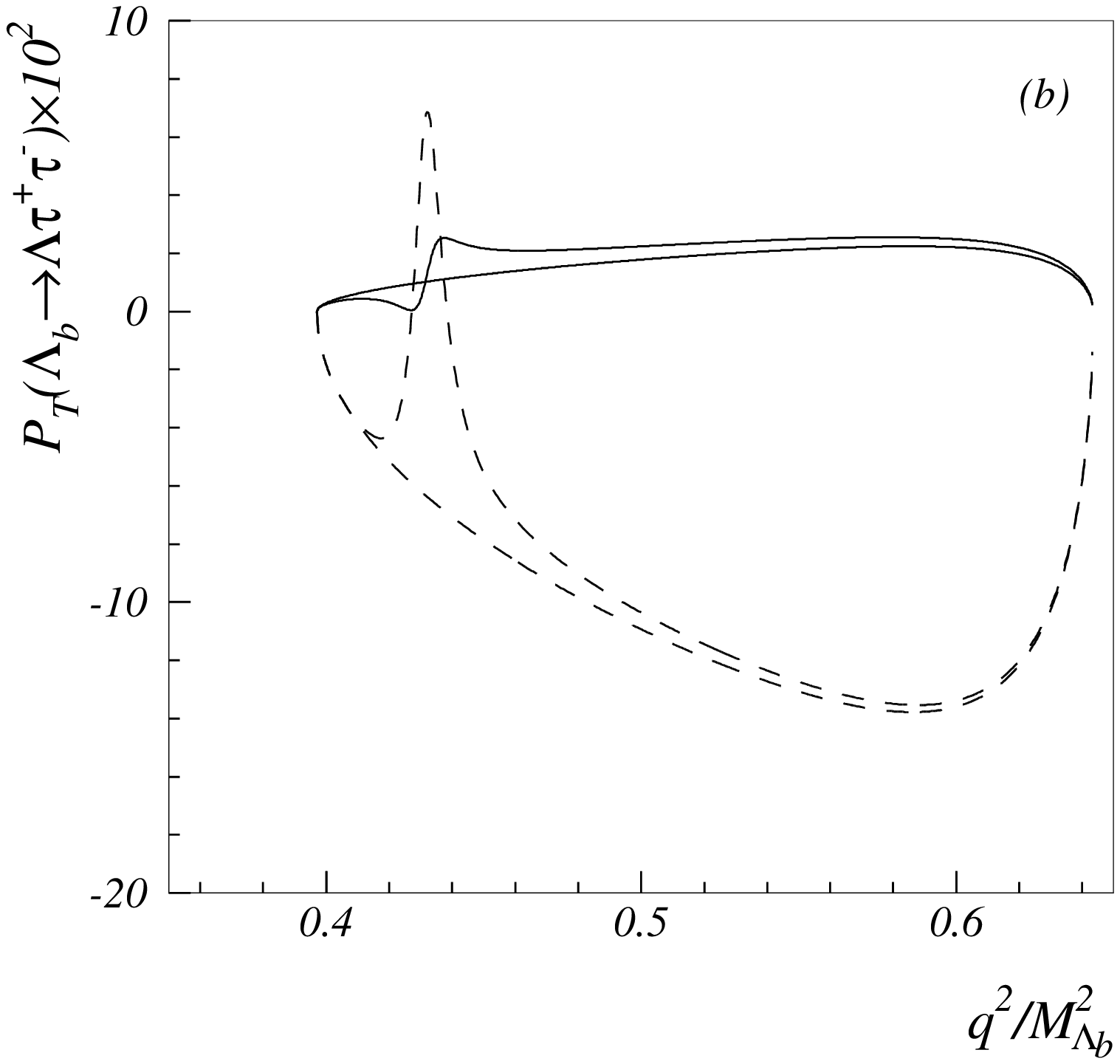}
\vskip 8.cm
\caption{
Same as Figure 5 but for the
transverse polarization asymmetries.}
\end{figure}

\newpage
\begin{figure}[h]
\includegraphics{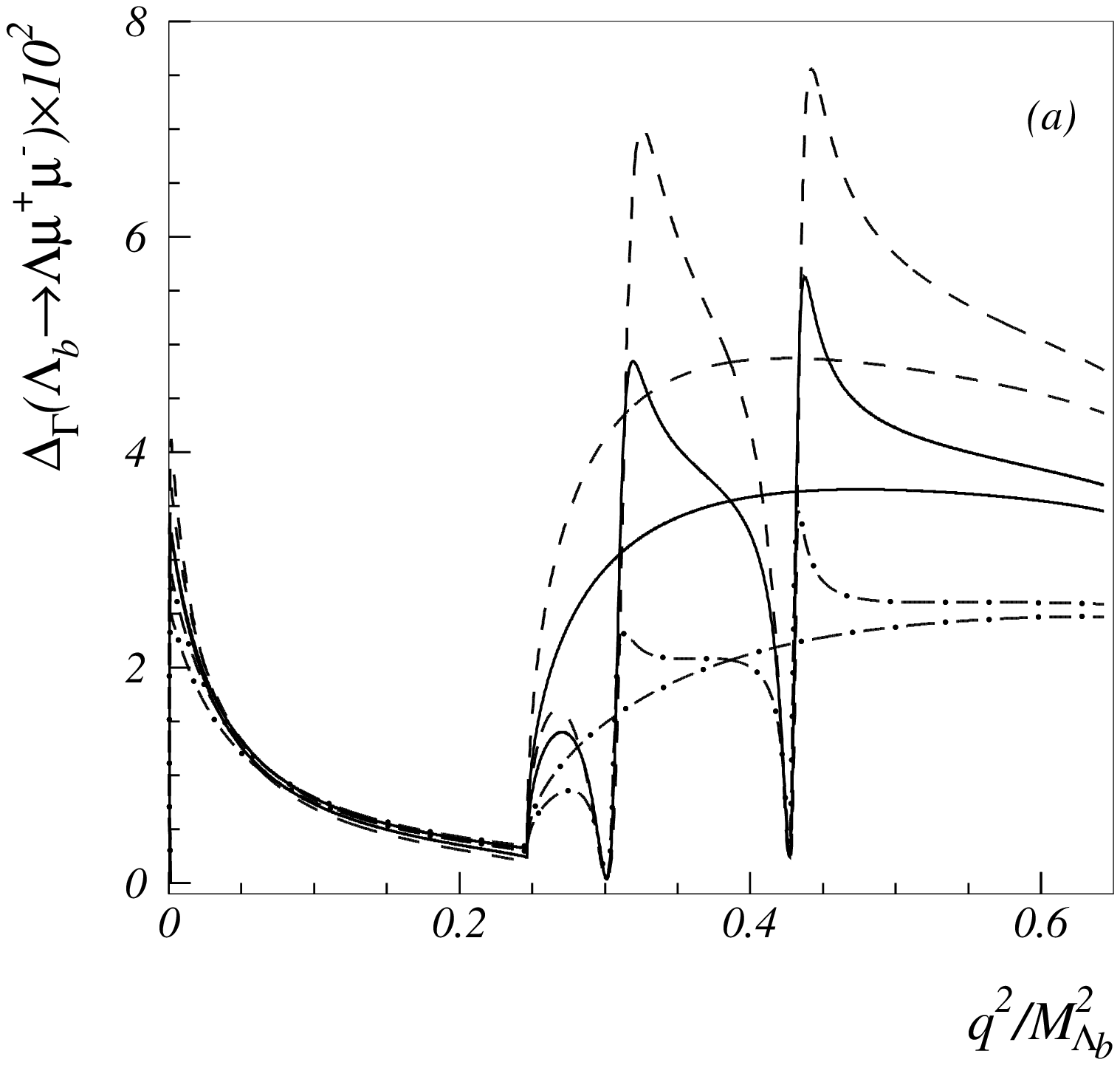}
\vskip 5.5cm
\end{figure}

\vskip 2.cm
\begin{figure}[h]
\includegraphics{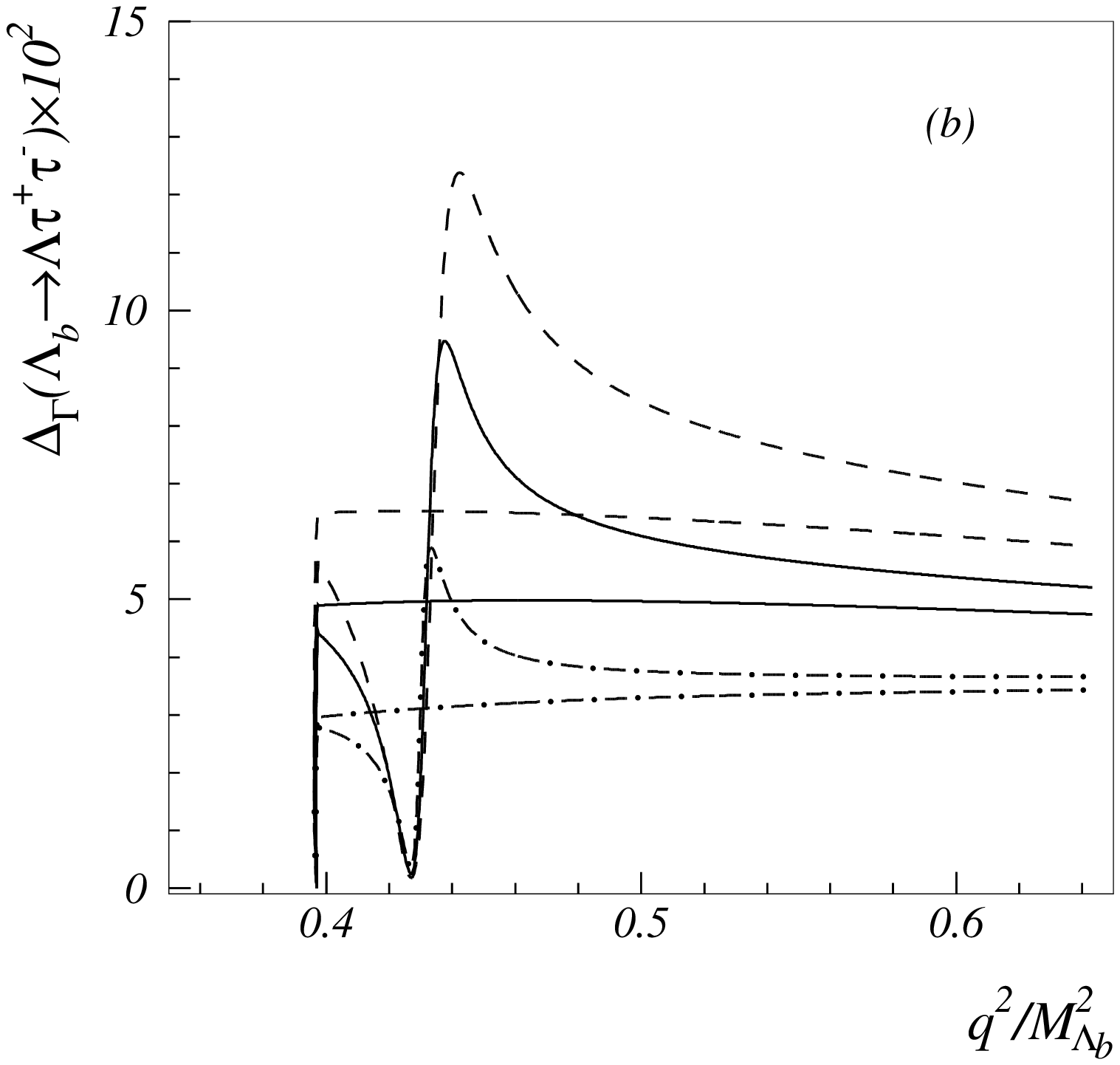}
\vskip 8.cm
\caption{
Same as Figure 5 but for
$\Delta_\Gamma$.}
\end{figure}

\newpage
\begin{figure}[h]
\includegraphics{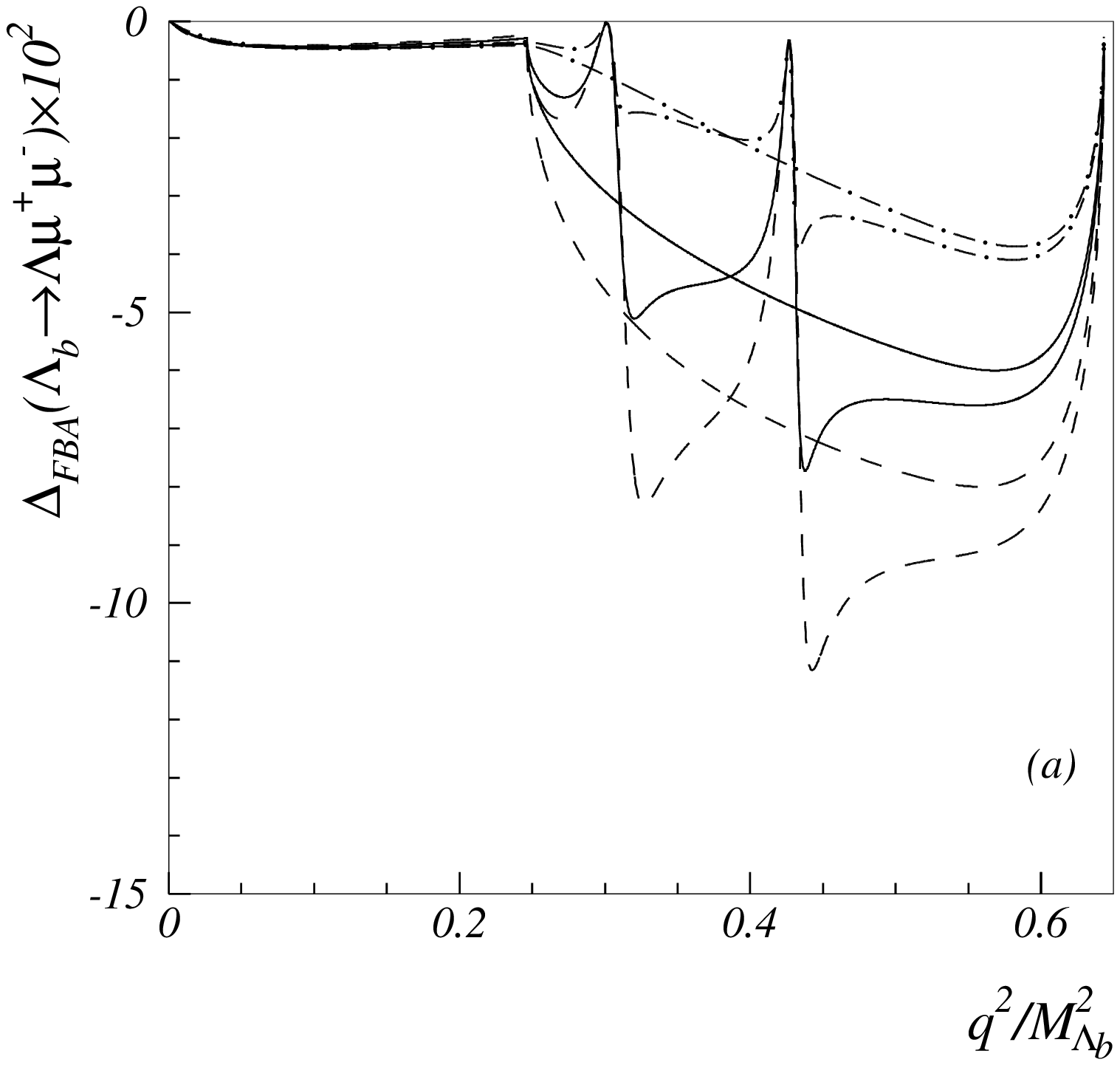}
\vskip 5.5cm
\end{figure}

\vskip 2.cm
\begin{figure}[h]
\includegraphics{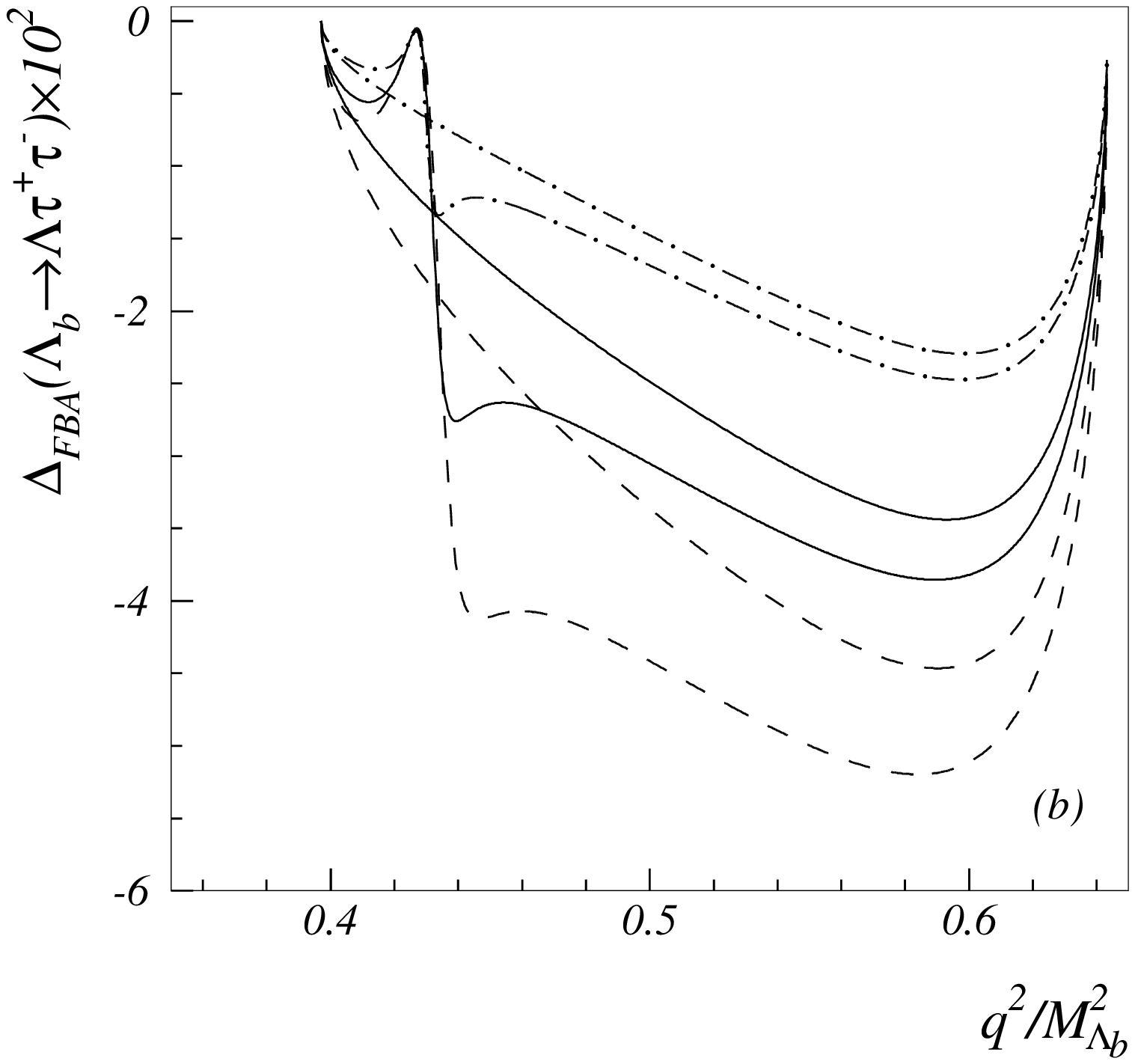}
\vskip 8.cm
\caption{
Same as Figure 5 but for
$\Delta_{FBA}$.}
\end{figure}

\newpage
\begin{figure}[h]
\includegraphics{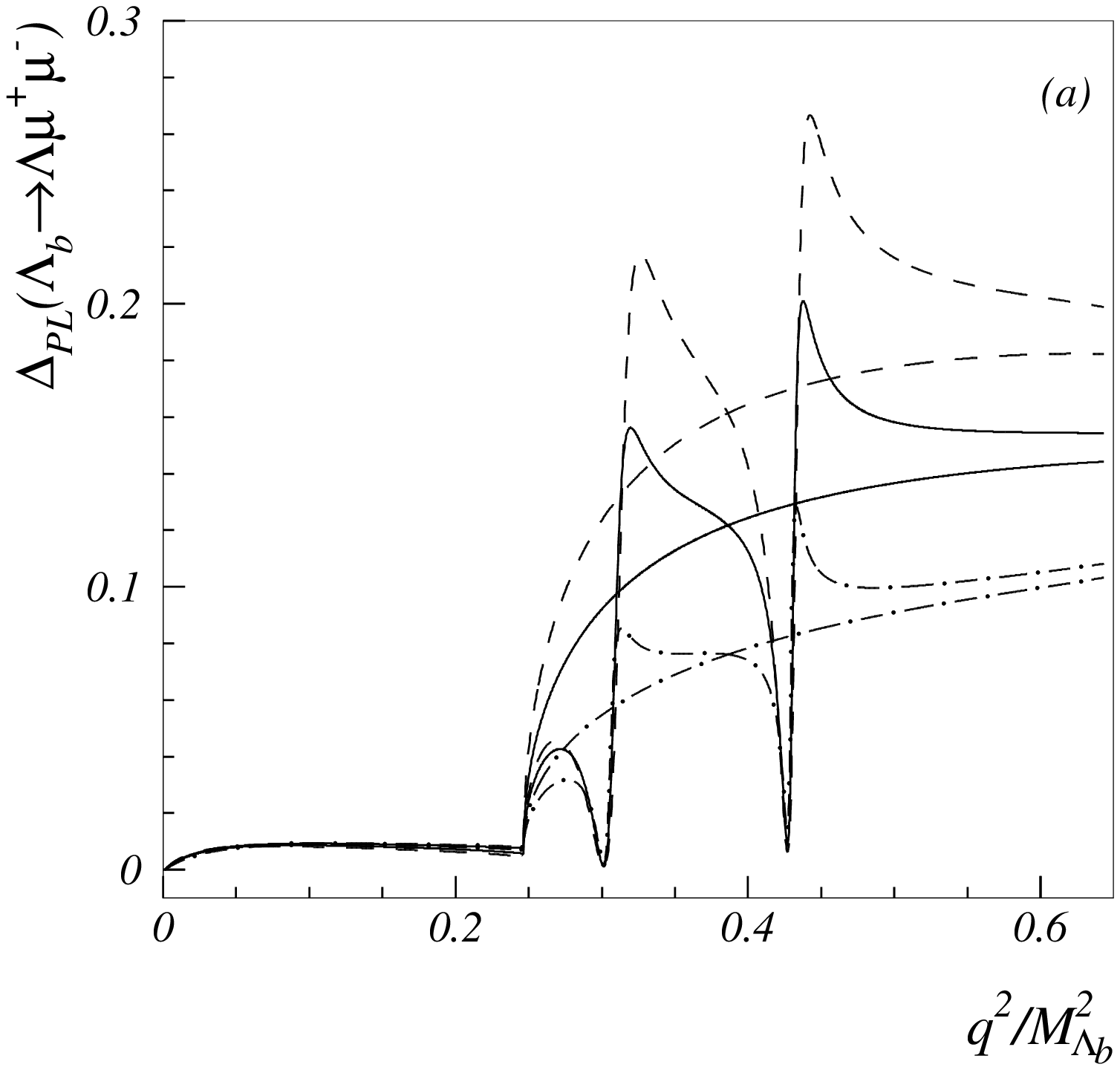} \vskip 5.5cm
\end{figure}

\vskip 2.cm
\begin{figure}[h]
\includegraphics{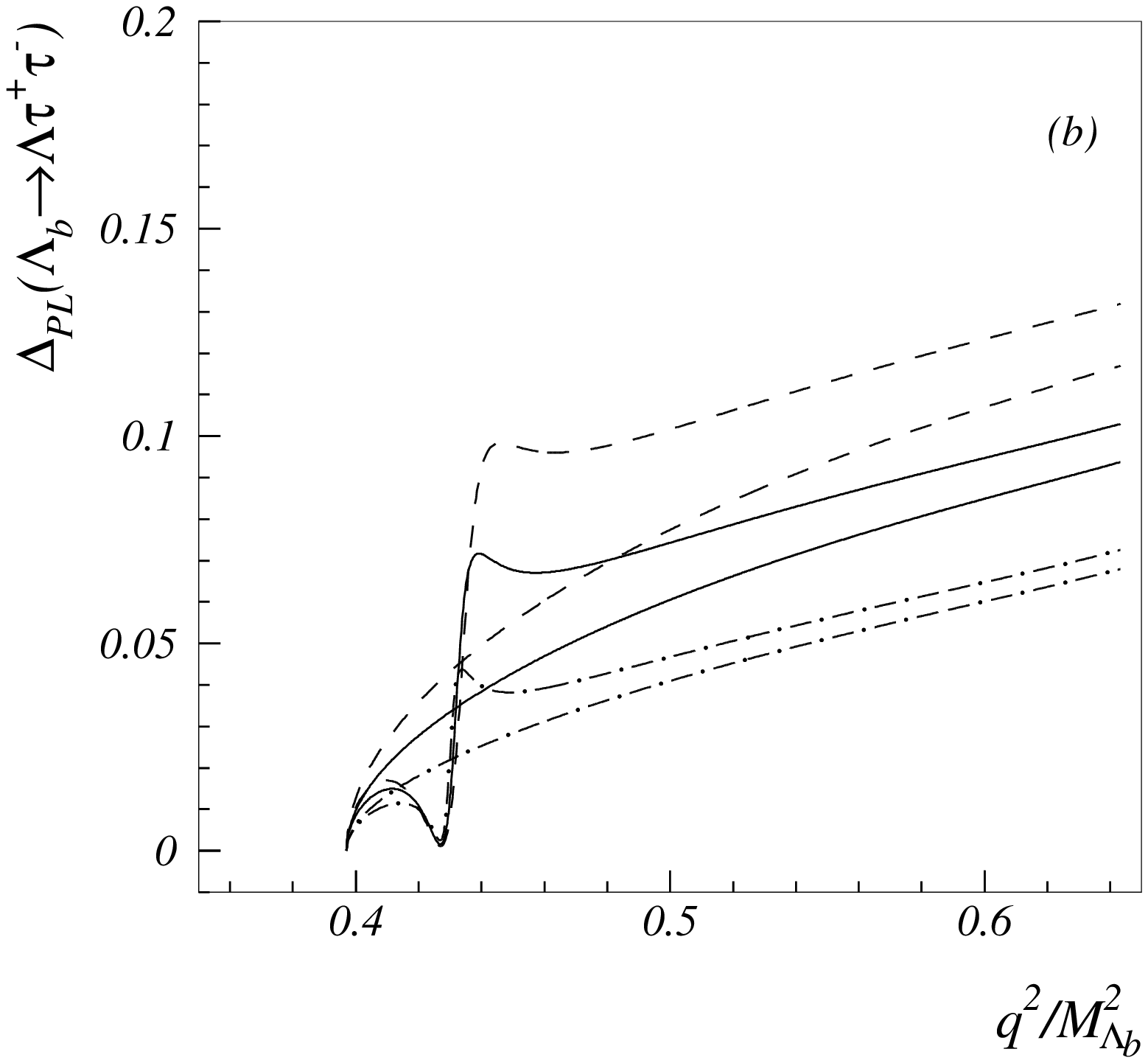} \vskip 8.cm
\caption{
Same as Figure 5 but for
$\Delta_{P_L}$.}
\end{figure}

\newpage
\begin{figure}[h]
\includegraphics{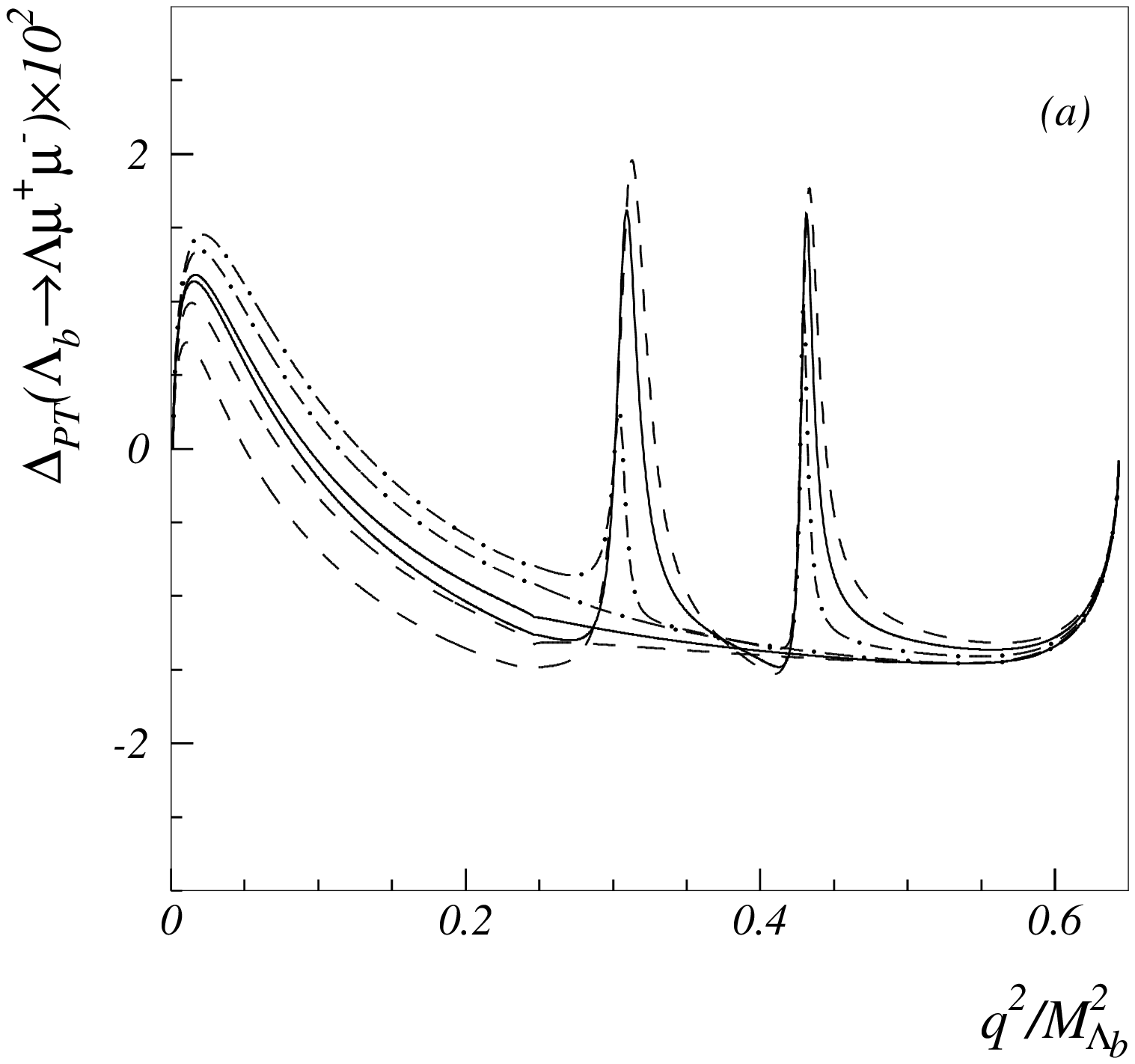} \vskip 5.5cm
\end{figure}

\vskip 2.cm
\begin{figure}[h]
\includegraphics{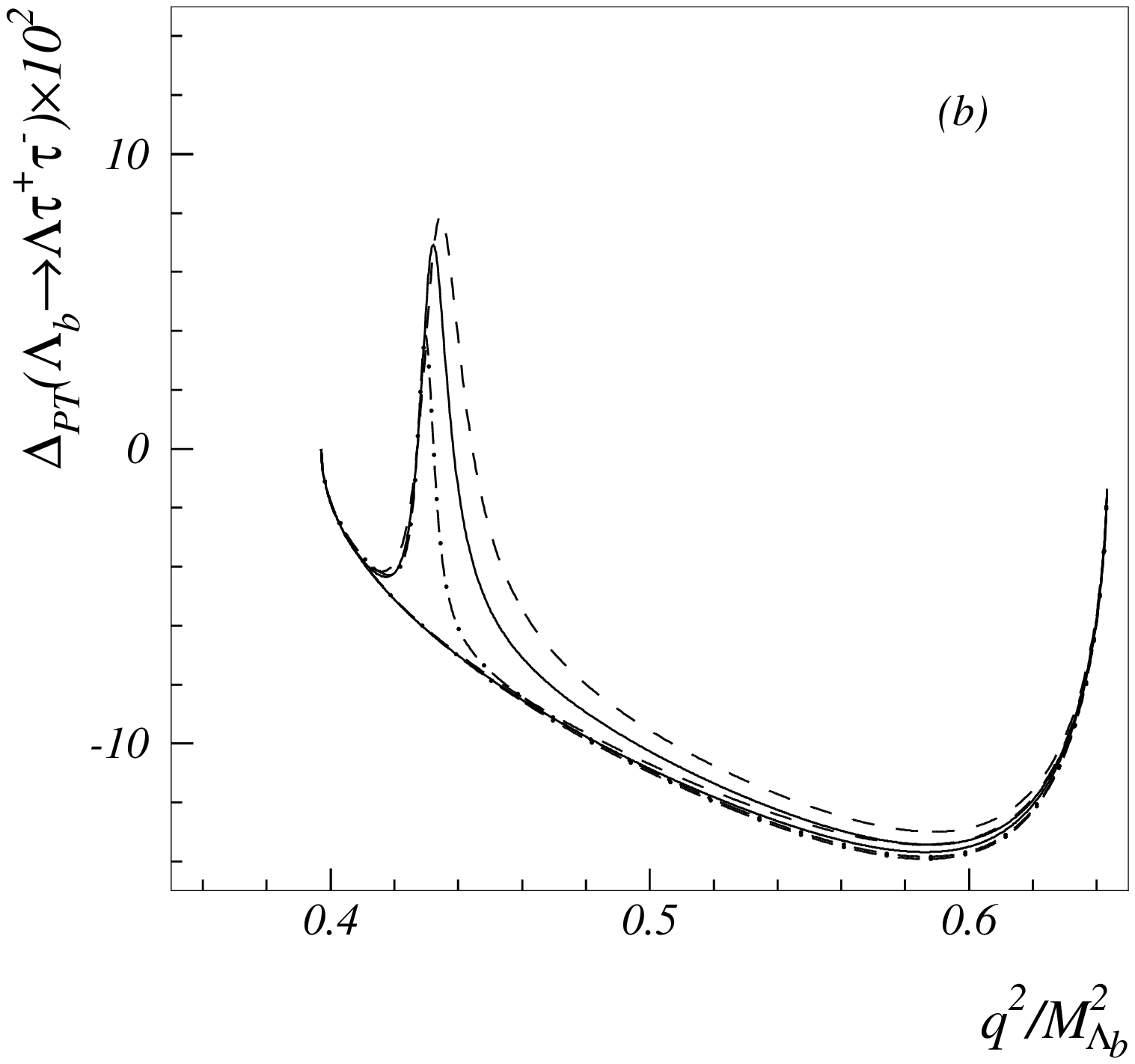} \vskip 8.cm
\caption{
Same as Figure 5 but for
$\Delta_{P_T}$.}
\end{figure}

\end{document}